\documentclass[fleqn,usenatbib]{mnras}

\usepackage{newtxtext,newtxmath}

\usepackage[T1]{fontenc}
\usepackage{ae,aecompl}

\usepackage{graphicx}	
\usepackage{amsmath}	
\usepackage{amssymb}	
\usepackage{multirow}
\usepackage{booktabs}

  \newcommand{\mbh}{$M_\mathrm{BH}$}

  \newcommand{\mdot}{$\dot{m}$}

  \newcommand{\xmm}{{\it XMM-Newton}}
  
  \newcommand{\sxone}{$\left(\frac{F_{\mathrm{bb}}}{F_{\mathrm{bmc}}}\right)_{0.5-2\ \mathrm{keV}}$}
  \newcommand{\sxtwo}{$\left(\frac{F_{\mathrm{bb}}}{F_{\mathrm{bmc}}}\right)_{0.5-10\ \mathrm{keV}}$}
  \newcommand{\sxthree}{$({L_{\mathrm{bb,0.5-2\ \mathrm{keV}}}}/{L_{\mathrm{Edd}}})$}
  \newcommand{\xspec}{\textsc{xspec}} 

\title[The soft X-ray excess: NLS1s vs. BLS1s]{The soft X-ray  excess: NLS1s vs. BLS1s}

\author[Gliozzi \& Williams]{
Mario Gliozzi$^{1}$\thanks{E-mail: mgliozzi@gmu.edu}
and James K. Williams$^{1}$
\\
$^{1}$ Department of Physics and Astronomy,
George Mason University, 4400 University Drive, Fairfax, VA 22030\\
}

\date{Accepted XXX. Received YYY; in original form ZZZ}

\pubyear{2015}

\begin{document}
\label{firstpage}
\pagerange{\pageref{firstpage}--\pageref{lastpage}}
\maketitle

\begin{abstract}
The soft X-ray excess -- the excess of X-rays below 2 keV with respect to the extrapolation of the hard X-ray spectral continuum model -- is a very common feature among type 1 active galactic nuclei (AGN); yet the nature of the soft X-ray excess is still poorly understood and hotly debated. To shed some light on this issue, we have measured in a model-independent way the soft excess strength in a flux-limited sample of broad-line and narrow-line Seyfert 1 galaxies (BLS1s and NLS1s) that are matched in X-ray luminosity but different in terms of the black hole mass and the accretion rate values, with NLS1s being characterized by smaller $M_{\rm BH}$ and larger $\dot m$ values. Our analysis, in agreement with previous studies carried out with different AGN samples, indicates that: 1) a soft excess is ubiquitously detected in both BLS1s and NLS1s; 2) the strength of the soft excess is significantly larger in the NLS1 sample, compared to the BLS1 sample; 3) combining the two samples, the strength of the soft excess appears to positively correlate with the photon index as well as with the accretion rate, whereas there is no correlation with the black hole mass. Importantly, our work also reveals the lack of an anticorrelation between the soft excess strength and the luminosity of the primary X-ray component, predicted by the absorption and reflection scenarios. Our findings suggest that the soft excess is consistent with being produced by a warm Comptonization component. Larger, more complete samples of NLS1s and BLS1s are needed to confirm these conclusions.
\end{abstract}




\section{Introduction}
In the current standard model, active galactic nuclei (AGN) are powered by accretion onto supermassive black holes, which produces the most powerful stationary sources in the universe with radiation emitted throughout the electromagnetic spectrum \citep[e.g.,][]{frank92}. The X-ray portion of the spectrum provides crucial information about the central engines of AGN, because X-rays are produced in the innermost part of the accretion flow, are less affected by absorption, and are easier to disentangle from the stellar and galactic contributions, compared to optical and UV. It is now widely accepted that the primary X-ray emission from 2 to hundreds of keV, generally parametrized by a power-law spectral component, is produced via Comptonization of  optical/UV seed  photons from a standard accretion disk in a hot corona  \citep[e.g.,][]{mush80,haar91}, and is then modified by neutral or ionized absorption, and by reflection from the accretion disk itself \citep[e.g.,][]{zdzi90} or from a more distant medium, such as the putative torus \citep[e.g.,][]{ghis94}.

If the primary X-ray continuum model is extrapolated to lower energies, the soft X-ray data generally lie well above the model, yielding the so-called soft X-ray excess. Despite the fact that the soft excess has been observed for several decades \citep[e.g.,][]{prav81,sing85}, and is nearly ubiquitous in type 1 AGN \citep[e.g.,][]{walt93,pico05}, its nature is still hotly debated. Over the years, different hypotheses have been proposed to explain the soft excess, such as thermal emission from the disk \citep{turn89}, relativistically blurred reflection \citep{ball01}, relativistically smeared absorption \citep{gierl04},  or a warm Comptonized component \citep[e.g.,][]{magdz98}. Although these models describe very different physical scenarios, they are often able to fit the same spectral data equally well \citep[e.g.,][]{gierl06,crum06}. To break this spectral degeneracy, the simultaneous spectral coverage of a broader energy band including optical, UV, and X-rays may be helpful \citep[e.g.,][]{mehdi11}. Additionally, model-independent constraints obtained from the temporal analysis may rule out some of the competing models at least for some specific sources \citep[e.g.,][]{edel96,emma11,glioz13}. 

An alternative approach to investigate the nature of the soft X-ray excess is based on the determination of its strength regardless of the specific spectral model used and on correlation analyses in sizable samples of AGN \citep[e.g.,][]{bian09a,bian09b,bois16}. This is the approach adopted in this work: we systematically analyze the strength of the soft excess of two samples of broad-line Seyfert 1 galaxies (BLS1s) and narrow-line Seyfert 1 galaxies (NLS1s), with matching X-ray luminosity distributions observed with \xmm. In paper I \citep{will18}, we have analyzed their 2--10 keV spectral properties and homogeneously estimated their black hole masses using an X-ray scaling method that is independent of the inclination angle and unaffected by the putative varying contributions of the radiation pressure.

In this work we extend our spectral analysis to the 0.5--10 keV energy range to investigate the nature of the soft excess. Specifically we try to address the following outstanding questions: Is there any difference in the soft excess properties between BLS1s and NLS1s? Is there any correlation between the soft excess strength and the fundamental parameters of these BH systems, such as  \mbh\ and \mdot?  Does our sample (and its BLS1 and NLS1 subsamples) show a positive correlation between the 2--10 keV photon index $\Gamma$ and accretion rate?

The paper is structured as follows. In Section 2, we briefly describe the data reduction and the properties of the sample, summarizing the main findings from paper I. In Section 3, we perform a systematic spectral analysis of our sample, and in Section 4 we constrain and compare the soft excess strength of BLS1s and NLS1s. Section 5 deals with the correlation analysis, whereas in Section 6 we discuss the main findings, compare them with the literature, and draw our conclusions.

Hereafter, we adopt a cosmology with $H_0=71{\rm~km~s^{-1}~Mpc^{-1}}$,
$\Omega_\Lambda=0.73$ and $\Omega_{\rm M}=0.27$ \citep{ben03}. 

\section{Sample and Data Reduction}
Our sample was derived from the flux-limited sample of \citet{zhou10} of type 1 AGN observed with \xmm\ with $f_{\rm 2-10 ~keV} \ge 10^{-12}~{\rm erg ~s^{ - 1} ~cm^{ - 2}}$. This sample broadly overlaps with the CAIXA catalog \citep{bian09a} with 86 common objects. As explained in paper I, we performed the data reduction following the standard procedures of Science Analysis System (SAS) version 15.0.0 and systematically analyzed the spectral properties of 98 objects out of the original 114 AGN, after excluding the narrow emission-line galaxies. We constrained the  \mbh\ of 89 AGN using the X-ray scaling method, which is  described in detail in \citet{glioz11}. After a closer inspection, we realized that five objects with FWHM H$\beta$  $> 2000~{\rm km~s^{-1}}$ were erroneously classified as NLS1s in the sample of \citet{zhou10}  and consequently in paper I. After this correction, our sample contains 30 NLS1s and 59 BLS1s, whose redshift, Galactic absorption, \xmm\ observation identification number, and net exposure are reported in Tables~\ref{tab:1} and ~\ref{tab:2}. The Galactic $N_{\rm H}$ values, reported in these tables,
were obtained from NASA's HEASARC nH column density tool that uses the Leiden/Argentine/Bonn map, setting the cone radius to 0.5 degrees, and taking the weighted average \citep{kalber05}.

\begin{table*}
	\caption{NLS1 sample}
	\begin{center}
		\begin{tabular}{lcccc} 
			\toprule
			\toprule       
			Name & $z$ &  $N_\mathrm{H,Gal}$ & \xmm & Net exposure \\
			& & $(10^{20}\ \mathrm{cm}^{-2})$ & ObsID & (ks) \\
			\midrule
			I Zw 1 & 0.05890 & 4.61 & 0110890301 & $18.6$ \\
			Ton S180 & 0.06198 & 1.28 & 0764170101 & $88.4$ \\
			Mrk 359 & 0.01739 & 4.17 & 0655590501 & $13.9$ \\
			Mrk 1014 & 0.16311 & 2.28 & 0101640201 & $6.0$ \\
			Mrk 586 & 0.15554 & 2.73 & 0048740101 & $17.2$ \\
			Mrk 1044 & 0.01645 & 3.26 & 0695290101 & $59.1$ \\
			RBS 416 & 0.07100 & 1.19 & 0140190101 & $24.9$ \\
			HE 0450-2958 & 0.24657 & 1.82 & 0153100101 & $11.4$ \\
			PKS 0558-504 & 0.13720 & 3.36 & 0555170601 & $74.7$ \\
			Mrk 110 & 0.03529 & 1.33 & 0201130501 & $32.8$ \\
			RE J1034+396 & 0.04244 & 1.28 & 0506440101 & $76.7$ \\
			PG 1211+143 & 0.08090 & 3.06 & 0745110701 & $89.2$ \\
			PG 1244+026 & 0.04818 & 1.98 & 0744440501 & $79.2$ \\
			IRAS 13349+2438 & 0.10764 & 1.05 & 0402080301 & $41.9$ \\
			PG 1402+261 & 0.16400 & 1.36 & 0400200101 & $20.0$ \\
			PG 1440+356 & 0.07906 & 1.00 & 0005010301 & $18.2$ \\
			Mrk 493 & 0.03133 & 2.14 & 0744290101 & $50.0$ \\
			Mrk 896 & 0.02642 & 3.25 & 0112600501 & $7.1$ \\
			Mrk 1513 & 0.06298 & 3.70 & 0150470701 & $25.3$ \\
			II Zw 177 & 0.08135 & 4.80 & 0103861201 & $8.6$ \\
			Ark 564 & 0.02468 & 5.49 & 0670130901 & $37.8$ \\
			AM 2354-304 & 0.03029 & 1.52 & 0103861501 & $3.6$ \\
			\bottomrule
		\end{tabular}
	\end{center}
	\label{tab:1}
\end{table*}  

\begin{table*}
	\caption{BLS1 sample}
	\begin{center}
		\begin{tabular}{lcccc} 
			\toprule
			\toprule       
			Name & $z$ &  $N_\mathrm{H,Gal}$ & \xmm & Net exposure \\
			& & $(10^{20}\ \mathrm{cm}^{-2})$ & ObsID & (ks) \\
			\midrule
			PG 0052+251 & 0.15445 & 4.35 & 0301450401 & $6.4$ \\
			Q 0056-363 & 0.16414 & 1.81 & 0401930101 & $41.0$ \\
			Mrk 1152 & 0.05271 & 1.84 & 0147920101 & $20.5$ \\
			ESO 244-G17 & 0.02350 & 2.13 & 0103860901 & $17.1$ \\
			Fairall 9 & 0.04702 & 3.08 & 0605800401 & $88.1$ \\
			Mrk 590 & 0.02638 & 2.70 & 0201020201 & $20.5$ \\
			ESO 198-G24 & 0.04550 & 2.95 & 0305370101 & $78.8$ \\
			Fairall 1116 & 0.05857 & 2.06 & 0301450301 & $14.0$ \\
			1H 0419-577 & 0.10400 & 1.18 & 0604720301 & $49.9$ \\
			3C 120 & 0.03301 & 10.3 & 0693781601 & $30.3$ \\
			H 0439-272 & 0.08350 & 2.74 & 0301450101 & $13.5$ \\
			MCG-01-13-25 & 0.01589 & 3.73 & 0103863001 & $3.6$ \\
			MCG-02-14-09 & 0.02845 & 9.49 & 0550640101 & $79.9$ \\
			MCG+08-11-11 & 0.02048 & 18.9 & 0201930201 & $18.5$ \\
			PMN J0623-6436 & 0.12889 & 3.87 & 0103860101 & $5.3$ \\
			ESO 209-G12 & 0.04050 & 18.8 & 0401790301 & $6.0$ \\
			PG 0804+761 & 0.10000 & 3.14 & 0605110101 & $14.6$ \\
			MCG+04-22-42 & 0.03235 & 3.02 & 0312191401 & $6.4$ \\
			PG 0947+396 & 0.20590 & 1.66 & 0111290101 & $17.6$ \\
			PG 0953+414 & 0.23410 & 1.24 & 0111290201 & $11.2$ \\
			HE 1029-1401 & 0.08582 & 5.66 & 0203770101 & $24.2$ \\
			PG 1048+342 & 0.16701 & 1.63 & 0109080701 & $25.7$ \\
			PG 1115+407 & 0.15434 & 1.43 & 0111290301 & $15.0$ \\
			PG 1116+215 & 0.17650 & 1.33 & 0554380301 & $50.6$ \\
			HE 1143-1810 & 0.03295 & 3.14 & 0201130201 & $21.7$ \\
			PG 1202+281 & 0.16530 & 1.75 & 0109080101 & $12.7$ \\
			Mrk 205 & 0.07085 & 2.93 & 0401240501 & $45.4$ \\
			NGC 4593 & 0.00900 & 1.56 & 0740920601 & $18.7$ \\
			PG 1307+085 & 0.15500 & 2.26 & 0110950401 & $10.6$ \\
			PG 1322+659 & 0.16800 & 1.75 & 0109080301 & $8.2$ \\
			4U 1344-60 & 0.01288 & 114 & 0092140101 & $25.3$ \\
			Mrk 279 & 0.03045 & 1.56 & 0302480601 & $16.9$ \\
			PG 1352+183 & 0.15200 & 1.67 & 0109080401 & $8.6$ \\
			PG 1415+451 & 0.11358 & 0.719 & 0109080501 & $21.0$ \\
			PG 1416-129 & 0.12894 & 7.23 & 0203770201 & $23.6$ \\
			PG 1425+267 & 0.36382 & 1.50 & 0111290601 & $31.6$ \\
			PG 1427+480 & 0.22048 & 1.70 & 0109080901 & $34.7$ \\
			Mrk 841 & 0.03642 & 2.18 & 0763790501 & $19.9$ \\
			Mrk 290 & 0.02958 & 1.89 & 0400360801 & $13.2$ \\
			PG 1626+554 & 0.13300 & 1.45 & 0109081101 & $4.1$ \\
			PDS 456 & 0.18400 & 20.2 & 0721010601 & $110.6$ \\
			IGR J17418-1212 & 0.03700 & 20.2 & 0303230501 & $12.1$ \\
			Mrk 509 & 0.03440 & 4.29 & 0601391101 & $43.6$ \\
			MR 2251-178 & 0.06398 & 2.50 & 0763920801 & $21.7$ \\
			NGC 7469 & 0.01632 & 4.78 & 0760350801 & $58.9$ \\
			Mrk 926 & 0.04686 & 2.81 & 0109130701 & $6.9$ \\
			\bottomrule
		\end{tabular}
	\end{center}
	\label{tab:2}
\end{table*}

\begin{figure}
\includegraphics[width=0.99\columnwidth]{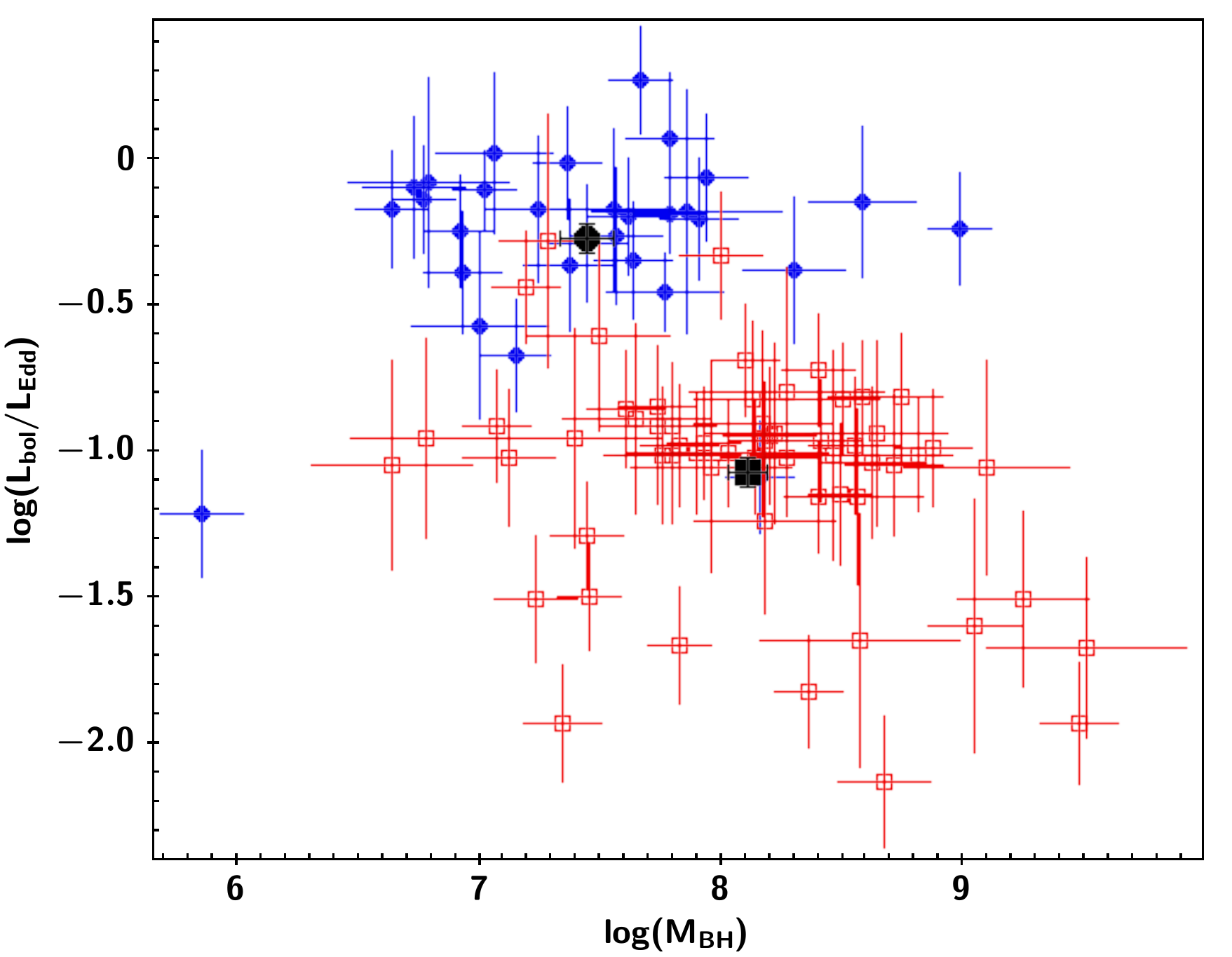}
\caption{Accretion rate values as measured by $\log(L_{\rm bol}/L_{\rm Edd)}$ plotted versus the logarithm of black hole masses $\log(M_{\rm BH})$. The open squares (red, in color) represent BLS1s, whereas the NLS1s are represented by filled circles (blue, in color). The same symbol convention is used throughout the paper. The larger black symbols indicate the respective mean values and the error bars represent their uncertainties.
}
\label{figure:fig1}
\end{figure}

After reclassifying five objects (PG~0953+414,PG~1115+407, PG~1116+215, PG~1322+659, and PDS~456) as BLS1s, we recomputed their photon indices, bolometric luminosities and the accretion rates in Eddington units, $\lambda_{\rm Edd}=L_{\rm bol}/L_{\rm Edd}$, using the appropriate bolometric corrections according to \citet{vasu09}. We then performed a statistical comparison between the distributions of $\Gamma$, \mbh, and $\lambda_{\rm Edd}$ for NLS1s and BLS1s. This reanalysis confirms the main findings of paper I at a higher significance level: NLS1s are characterized by steeper X-ray spectra with $\langle \Gamma \rangle_{\rm NLS1} =2.01\pm 0.05$ compared to $\langle \Gamma \rangle_{\rm BLS1} =1.72\pm 0.02$. NLS1s have smaller average BH masses, $\langle M_{\rm BH} \rangle_{\rm NLS1} =7.45\pm 0.12$, compared to BLS1s,  $\langle M_{\rm BH} \rangle_{\rm BLS1} =8.12\pm 0.08$, and their distributions are different according to a Kolmogorov-Smirnov test (with a chance probability of being drawn from the same population of $P_{\rm KS}=10^{-4}$) and a Student's $t$ test ($P_{\rm t}=10^{-5}$). Similarly, NLS1s and BLS1s differ very significantly in their accretion rate average values, $\langle \log\lambda_{\rm Edd} \rangle_{\rm NLS1} =-0.27\pm 0.06$ and $\langle \log\lambda_{\rm Edd} \rangle_{\rm BLS1} =-1.07\pm 0.05$, and in their distributions: $P_{\rm KS}=10^{-14}$ and  $P_{\rm t}=10^{-16}$. These differences are clearly illustrated in Figure~\ref{figure:fig1}, where the accretion rate, parametrized by $\log(L_{\rm bol}/L_{\rm Edd})$, is plotted versus $\log(M_{\rm BH})$. Despite a substantial overlap between the $\log(M_{\rm BH})$ distributions of BLS1s (indicated by open red squares) and NLS1s (blue filled circles), their average values (represented by the larger black symbols) are inconsistent with each other, whereas the  $\log(L_{\rm bol}/L_{\rm Edd})$  distributions are clearly separated with very little overlap.

\begin{figure*}
\includegraphics[bb=35 430 560 692,clip=,width=2.0\columnwidth]{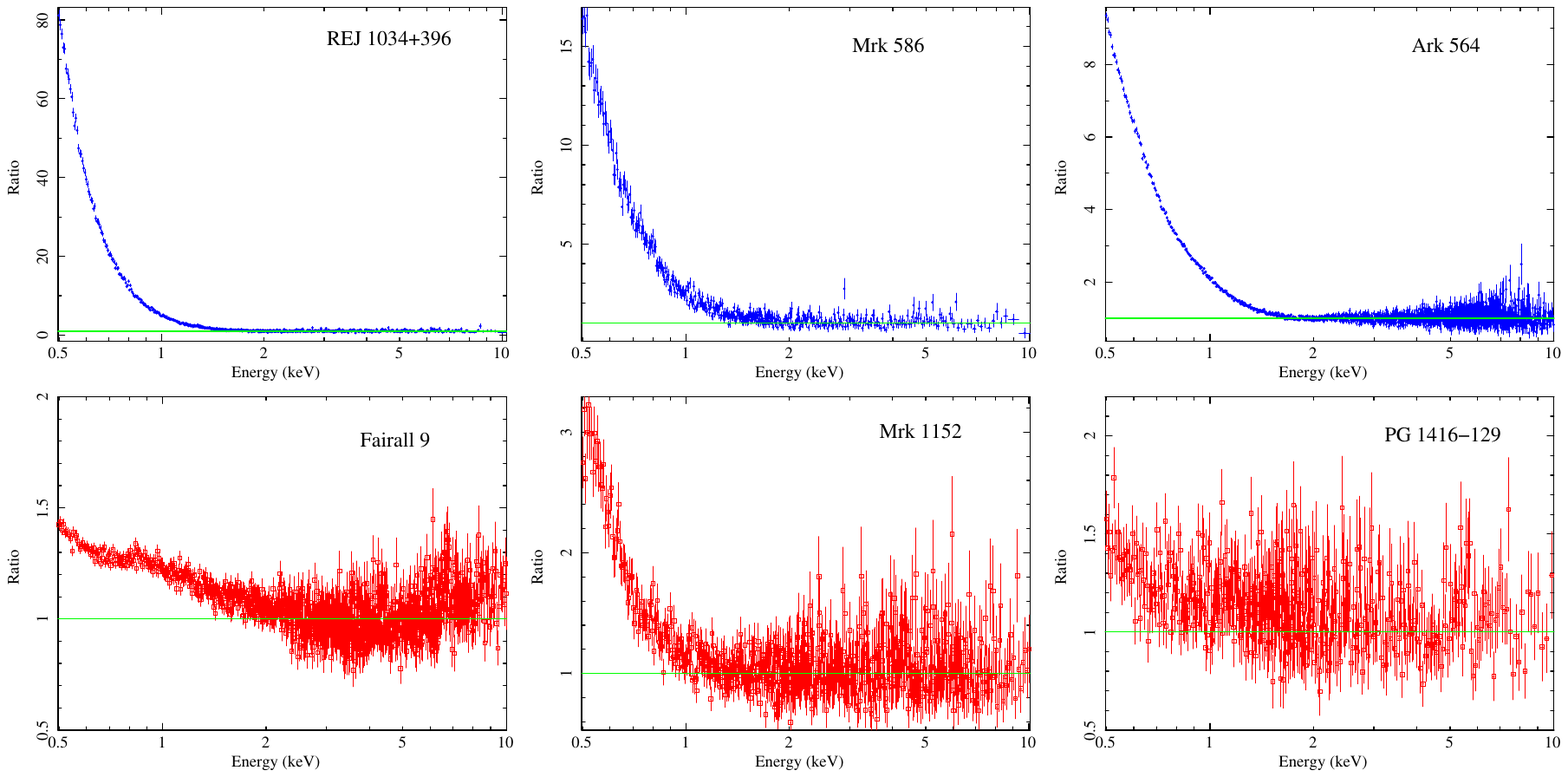}
\caption{Data-to-model ratio plots showing the soft excess for six  AGN:  the top row comprises the three objects with the strongest soft excess (which are all NLS1s, as indicated by the conventional symbol of blue filled circles), whereas the bottom row illustrates the three sources with the lowest soft excess (all BLS1s, indicated by red open squares). Note the diverse ranges on the y-axes of the different plots.}
\label{figure:fig2}
\end{figure*}
\section{Spectral analysis}
The X-ray spectral analysis  was performed using the \xspec\ {\tt v.12.9.0}
software package \citep{arn96}.  All spectra were re-binned within grppha 3.0.1 to have at least 20 counts per bin and then fitted using the  $\chi^2$ statistic. The errors on spectral parameters represent the 90\% confidence level. 

We carried out a systematic spectral analysis of every source starting with  the baseline model used for the 2--10  keV spectral analysis in paper I. This model comprises one Comptonization component, which represents the primary emission produced by the corona and is parametrized by the bulk motion Comptonization ({\tt BMC}) model in \xspec\ \citep{tita97}, and one Gaussian line ({\tt zgauss}) representing iron K\,$\alpha$ line emission, which is included only when required by the fit. To account for the Galactic and possible intrinsic local absorption, both additive components ({\tt BMC+zgauss}) are absorbed by a column density left free to vary with the minimum fixed at the Galactic value, and parametrized by the {\tt wabs} model in \xspec. 

When the baseline model is extrapolated down to 0.5 keV, the soft X-ray data lie considerably above the model, revealing the presence of soft excess in all 89 objects. The 100\% detection rate in both BLS1s and NLS1s confirms that this feature is ubiquitous in type 1 AGN. Examples of the soft excess in our sample are presented in Figure~\ref{figure:fig2}, which shows the data-to-model ratio plots of six objects illustrating the broad range associated with the soft excess: the top row shows the three objects with the strongest soft excess (with the strongest one, RE J1034+396, shown on the top left corner) and the bottom row shows the three objects with the weakest soft excess (with the weakest one, PG 1416-129, shown on the bottom right corner).

We kept the parameters of the Gaussian line and two parameters of the BMC model  
-- the temperature of the thermal photons and the Comptonization fration (which are less well constrained and do not affect the value of the flux) -- frozen to their best fit values obtained fitting the 2--10 keV range, whereas we left the BMC spectral index and normalization (i.e., the parameters that determine the flux) free to vary.

We then fitted the soft X-ray excess with a phenomenological model comprising one or two blackbody components. The new baseline model for the 0.5--10 keV energy range, expressed in the \xspec\ syntax, is:
\begin{verbatim}
wabs*(bbody+bbody+bmc+zgauss).
\end{verbatim}
Our baseline model does not include a Compton reflection component, because the \xmm\ energy range upper limit (10 keV) does not allow us to constrain it properly.

The results of this systematic spectral analysis of the 0.5--10 keV energy range are summarized in Tables~\ref{tab:3} and~\ref{tab:4} in the Appendix, which report the main spectral parameters (the temperature in keV for the blackbody components that parametrize the soft excess, and the photon index associated with the Comptonizing corona) with uncertainties as well as the reduced $\chi^2$ for each source. A few objects are not satisfactorily fitted with this baseline model. Including one or two components of partial covering absorption by partially ionized material, parametrized by {\tt zxipcf} in \xspec, makes the fits formally acceptable. Nevertheless, the presence of residuals in the soft part of the data-to-model ratio plots suggests that this modeling oversimplifies much more complex spectra. Indeed, a literature search reveals that nearly all the objects that cannot be fitted with our baseline model are characterized by the presence of several warm absorber components with different ionization states. Since the latter severely affect the spectra at low energies and hamper a proper characterization of the soft excess, we have excluded these sources (13 BLS1s and 8 NLS1s, which are reported in the Appendix) from further analysis. For the remainder of the paper, our ``clean'' sample comprises 46 BLS1s and 22 NLS1s.

\begin{figure}
\includegraphics[width=0.99\columnwidth]{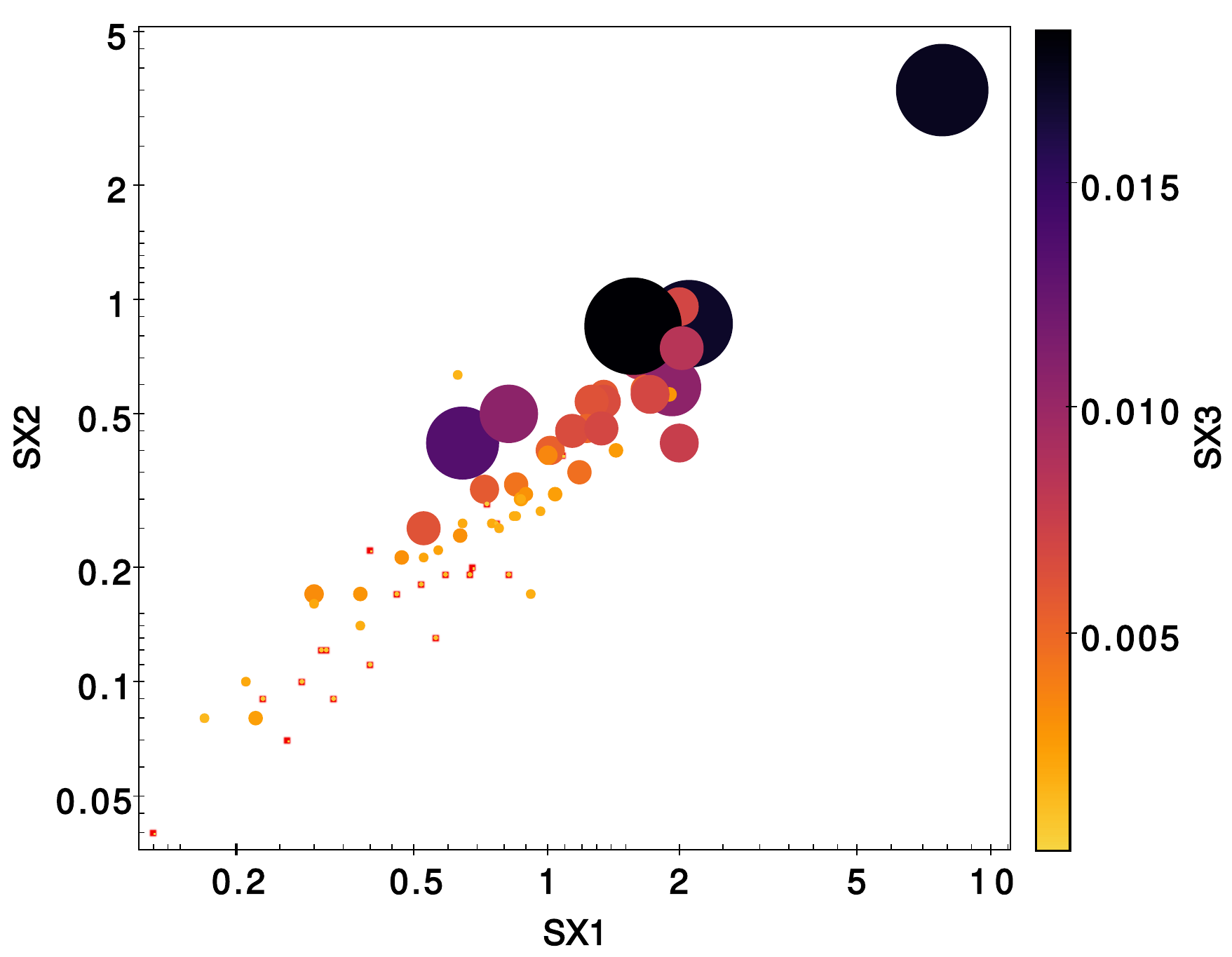}
\caption{Soft excess strengths visual comparison: $SX2$ is plotted vs. $SX1$ and the strength of $SX3$ is represented by the size and color of the symbols: the larger and the darker the symbols, the higher the values of $SX3$. The outlier in the top right corner is RE~J1034+396.}
\label{figure:fig3}
\end{figure}
\section{Soft Excess Strength}
Over the years, the strength of the soft X-ray excess has been quantified in different ways \citep[e.g.,][]{pico05,bian09a,petru13,bois16}. To allow a direct comparison with previous works and to make sure that our results do not depend on one specific measurement of the soft excess strength, we define three different quantities to characterize it in our sample.  

$SX1$ is the ratio of the unabsorbed 0.5--2 keV flux of the blackbody component over the analogous flux of the Comptonization component: 
$$
SX1=\left(\frac{F_{\mathrm{bb}}}{F_{\mathrm{bmc}}}\right)_{0.5-2\ \mathrm{keV}}.
$$

Similarly, $SX2$ is the ratio of the blackbody and Comptonization fluxes computed over the entire 0.5--10 keV band:
$$
SX2=\left(\frac{F_{\mathrm{bb}}}{F_{\mathrm{bmc}}}\right)_{0.5-10\ \mathrm{keV}}.
$$

Finally, $SX3$ is defined as the ratio of the 0.5--2 keV luminosity associated with the blackbody component and the Eddington luminosity:
$$
SX3=\left(\frac{L_{\mathrm{bb}_{0.5-2\ \mathrm{keV}}}}{L_{\mathrm{Edd}}}\right).
$$

We measured $SX1$, $SX2$, and $SX3$ for all objects of our clean sample and report their values as well as their uncertainties, obtained via error propagation, in Tables~\ref{tab:5} and~\ref{tab:6} in the Appendix. Note that the relative uncertainties of $SX3$ are considerably larger than those of $SX1$ and $SX2$, because they encompass the uncertainty in the \mbh\  estimation inherent in the X-ray scaling method, which directly affects the error of the Eddington luminosity.
\begin{figure*}
	\includegraphics[width=0.69\columnwidth]{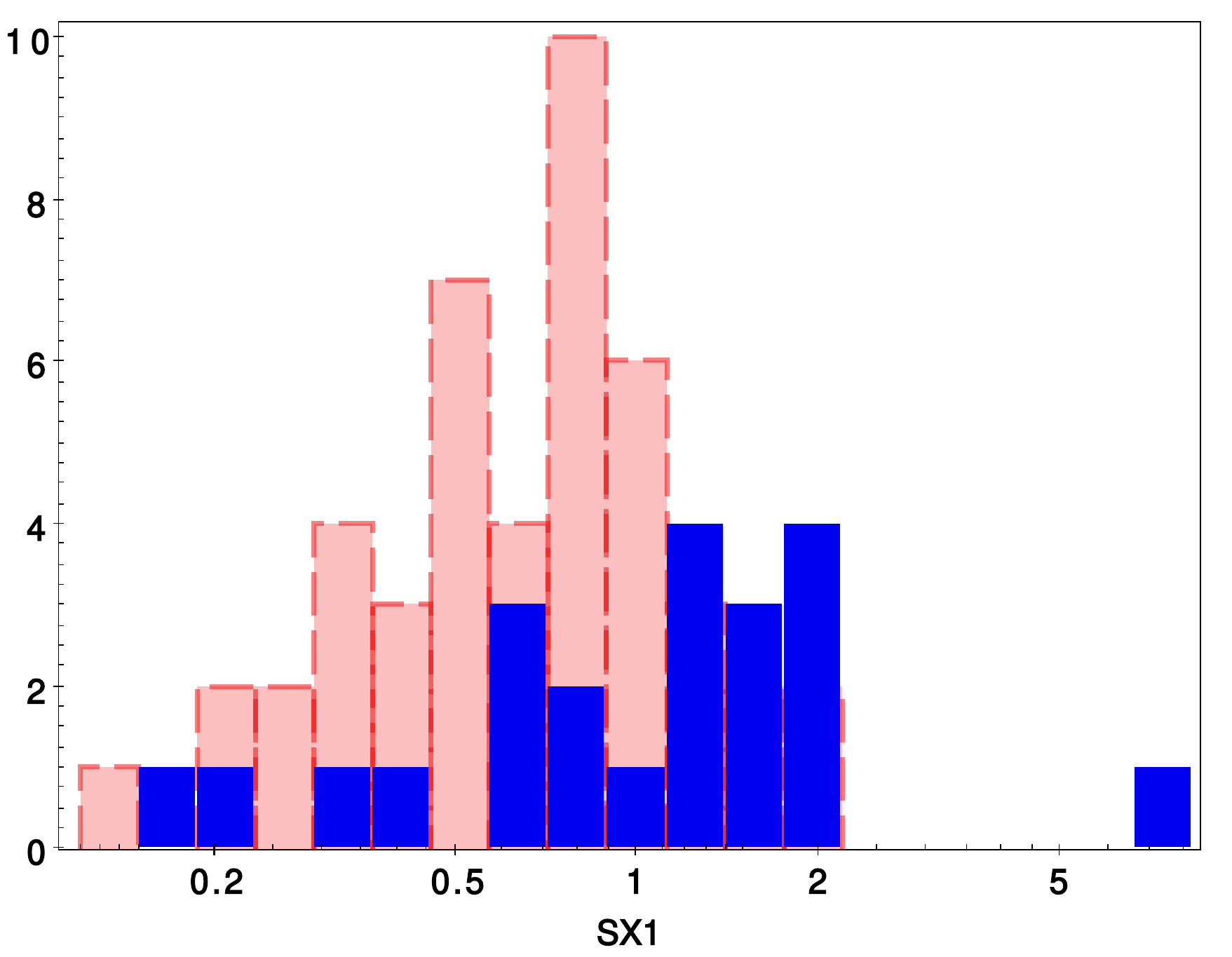}
	\hfill
	\includegraphics[width=0.69\columnwidth]{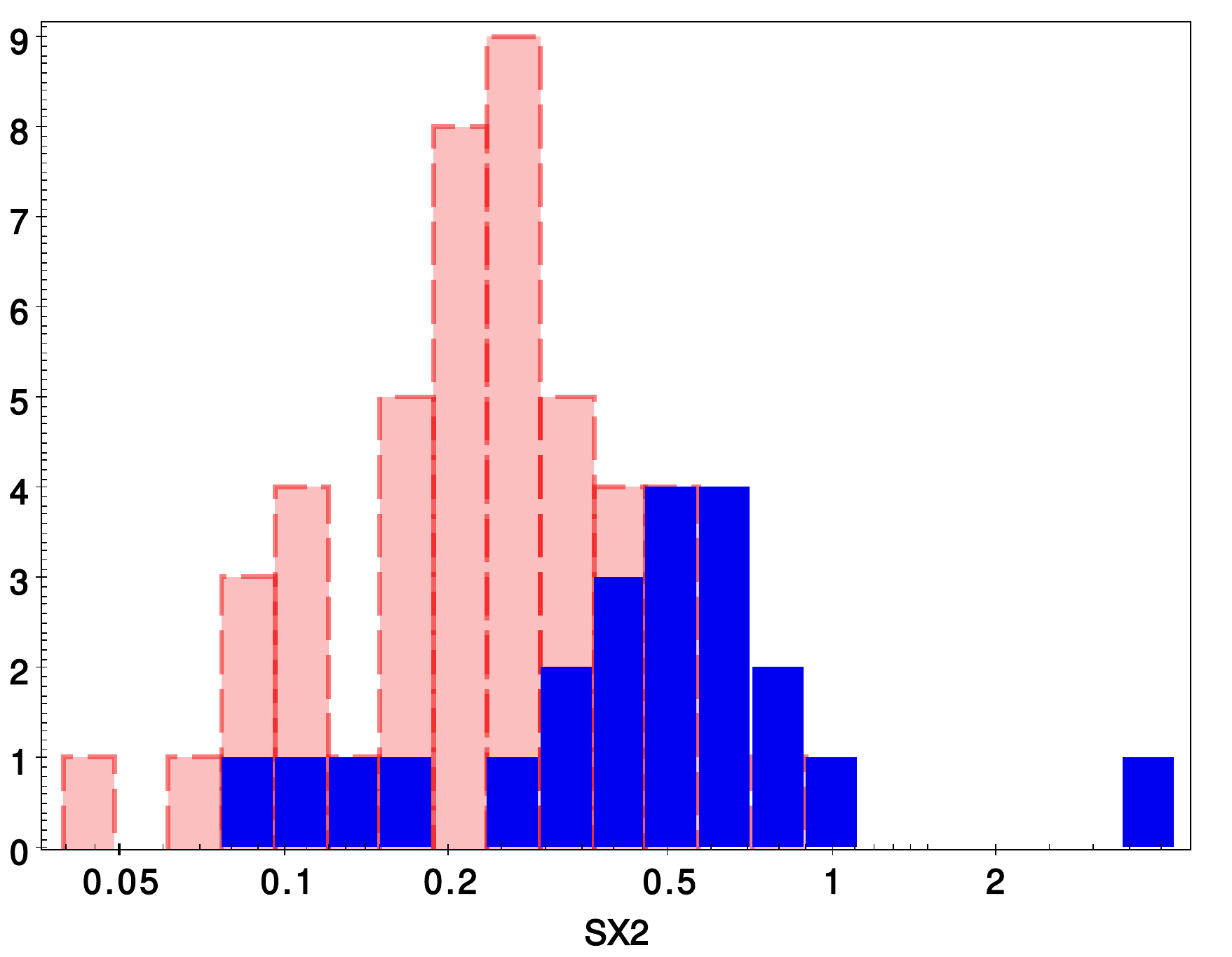}
	\hfill
	\includegraphics[width=0.69\columnwidth]{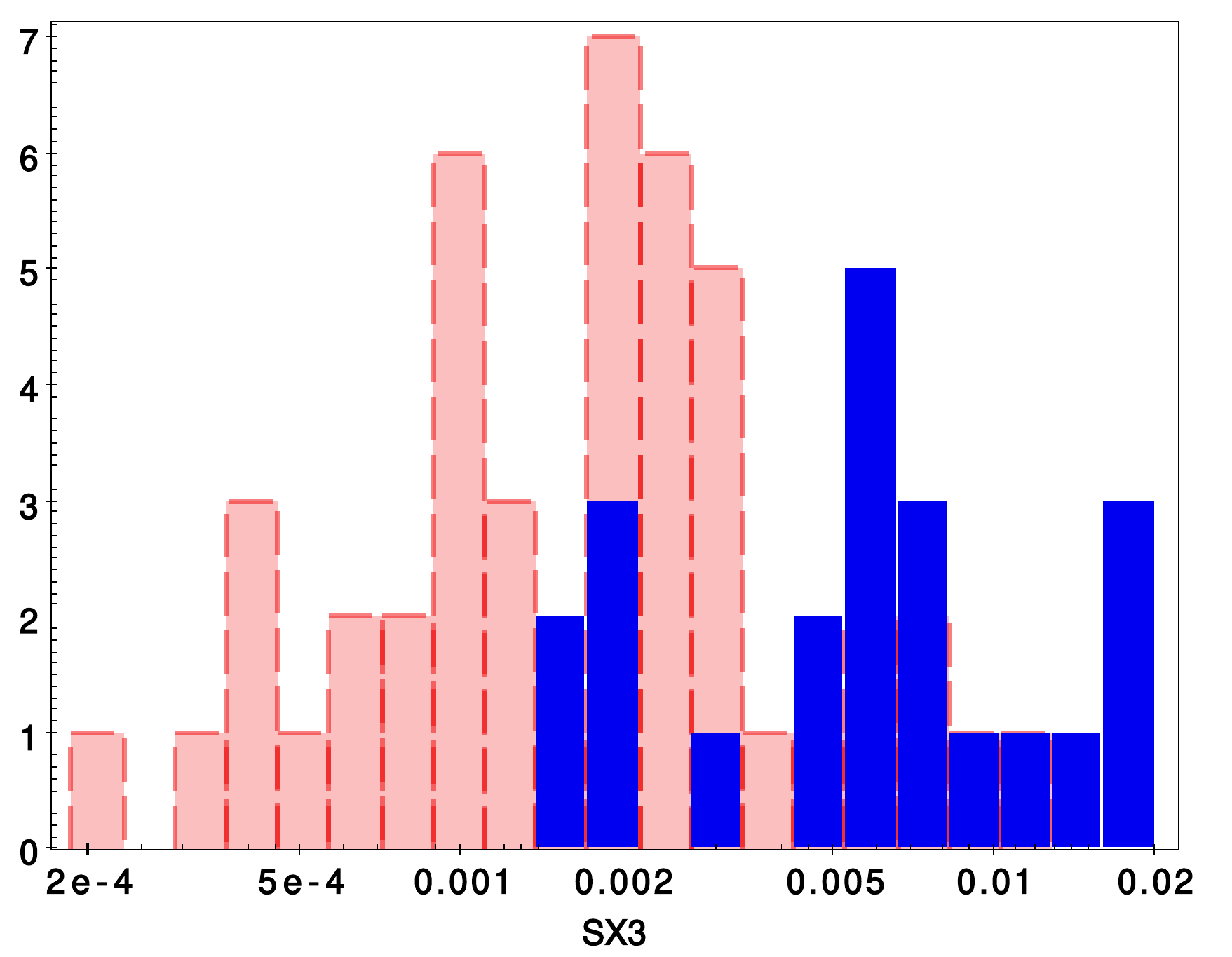}
	\caption{Histograms of the soft excess strength measurements $SX1$ (left panel), $SX2$ (middle panel), $SX3$ (right panel), for BLS1s, represented in shaded right color, and NLS1s in darker blue color.}
\label{figure:fig4}	
\end{figure*}

In Figure~\ref{figure:fig3} we plot  $SX2$ vs. $SX1$ and use the size and darkness of the color to illustrate the intensity of $SX3$: the larger and the darker the symbols, the higher the values of $SX3$. A visual inspection of Fig.~\ref{figure:fig3} clearly indicates that there is a strong correlation between $SX1$ and $SX2$, and that also $SX3$ appears to increase in concert with the first two measurements of the soft excess strength, since the darker and larger symbols are located in the top right part of the plot, whereas the smaller lighter color symbols are in the bottom left. These apparently strong correlations are formally confirmed by  Spearman's and Kendall's  rank correlation analyses. Specifically, for $SX2$ vs. $SX1$ $r=0.91~(P=10^{- 26})$ and $\tau=0.78~(P < 10^{- 26})$; for $SX3$ vs. $SX1$ $r=0.67~(P=10^{- 10})$ and $\tau=0.50~(P < 10^{- 26})$, and for  $SX3$ vs. $SX2$ $r=0.76~(P=10^{- 14})$ and $\tau=0.59~(P < 10^{- 26})$.

Figure~\ref{figure:fig4} illustrates the distributions of  $SX1$, $SX2$, and $SX3$ for BLS1s (shown in shaded red color) and NLS1s (blue filled histograms), suggesting that NLS1s have systematically larger values than BLS1s on average. This conclusion is quantitatively confirmed by K-S and Student's $t$ tests. NLS1s have $\langle SX1 \rangle_{\rm NLS1} =1.43\pm 0.32$ which is larger than $\langle SX1 \rangle_{\rm BLS1} =0.75\pm 0.06$, and their distributions are different according to a K-S test, with a chance probability of being drawn from the same population of $P_{\rm KS}=4\times 10^{-3}$ and a Student $t$  probability of $P_{\rm t}=7\times 10^{-3}$. Similarly, $\langle SX2 \rangle_{\rm NLS1} =0.61\pm 0.15$ and $\langle SX2 \rangle_{\rm BLS1} =0.26\pm 0.02$ with $P_{\rm KS}=10^{-4}$ and $P_{\rm t}=10^{-3}$. Even larger differences are obtained when comparing the distributions of $SX3$:  $\langle SX3 \rangle_{\rm NLS1} =(7.1\pm 1.1)\times 10^{- 3}$ and $\langle SX3 \rangle_{\rm BLS1} =(2.5\pm 0.3)\times 10^{- 3}$ with $P_{\rm KS}=2\times 10^{-5}$ and $P_{\rm t}=2\times 10^{-6}$. Note that the difference becomes even more significant (in terms of sigmas) if the NLS1 outlier (in $SX1$ and $SX2$) RE~J1034+396 is excluded from the statistical analysis.

\begin{figure}
\includegraphics[width=0.99\columnwidth]{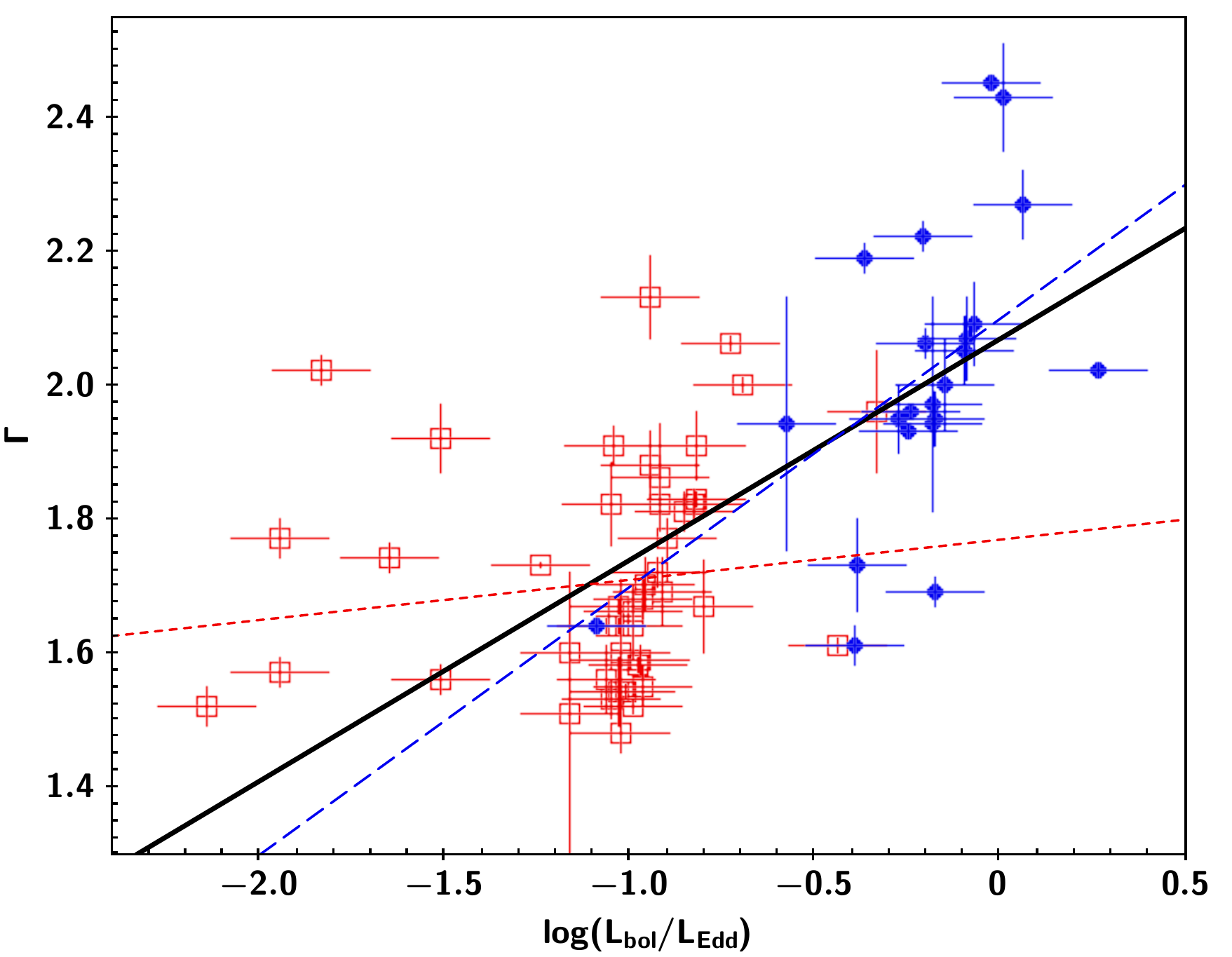}
\caption{$\Gamma$ vs. $ \log(L_{\rm bol}/L_{\rm Edd})$ for the clean sample. The continuous thick (black in color) line represents the linear regression obtained with the linmix$\_$err routine. The dotted (red in color) and the dashed (blue in color) lines represent the linear fit for the BLS1s and NLS1s sample, respectively.}
\label{figure:fig5}
\end{figure}
\section{Correlation analysis}
We carried out a systematic correlation analysis to investigate the existence of statistically significant positive or negative trends between relevant parameters and the different measurements of the soft excess strength. To this end, we have used linear regression routine {\tt linmix$\_$err} that performs the linear fit between two variables with errors on both x and y using a Bayesian approach \citep{kelly07}. For completeness, we also determined the linear fit using the \textsc{idl} routines {\tt ladfit},  which utilizes a robust least absolute deviation method, and {\tt linfit}, which minimizes the $\chi^2$ statistics taking into account the errors on the dependent variable only. Since the results of the three methods are broadly consistent with each other, in the text we will only quote the results from {\tt linmix$\_$err}. The significance of the linear correlations is further assessed using Spearman's and Kendall's rank correlation analyses.

\subsection{Photon index vs. accretion rate}
Before investigating the possible correlations between the soft excess strength and relevant physical quantities for BH systems, we examine the correlation between the photon index $\Gamma$ and the accretion rate as measured by  $\log(L_{\rm bol}/L_{\rm Edd})$. Our whole sample (comprising all BLS1s and NLS1s including also the sources with warm absorbers that were excluded from the soft excess analysis) shows a strong positive correlation, described by the equation $\Gamma = (2.09\pm 0.05) + (0.35\pm 0.05) \log(L_{\rm bol}/L_{\rm Edd})$, with  $r=0.63~(P=5\times 10^{- 11})$ and $\tau=0.48~(P < 10^{- 26})$. If the correlation analysis is run separately for NLS1s and BLS1s, positive trends with different slopes are found,  
$\Gamma_{\rm NLS1} = (2.19\pm 0.02) + (0.56\pm 0.05) \log(L_{\rm bol}/L_{\rm Edd)}$ and $\Gamma_{\rm BLS1} = (1.79\pm 0.01) + (0.08\pm 0.01) \log(L_{\rm bol}/L_{\rm Edd})$, but at a lower significance level: $r=0.50~(P=5\times 10^{- 3})$ and $\tau=0.38~(P =3\times10^{- 3})$ for NLS1s, and  $r=0.27~(P=3\times 10^{- 2})$ and $\tau=0.21~(P =1.5\times10^{- 2})$ for BLS1s. 

When we use the clean sample in this analysis, once more we find a strong, positive correlation, which is  described by the equation $\Gamma_{\rm clean} = (2.07\pm 0.05) + (0.33\pm 0.06) \log(L_{\rm bol}/L_{\rm Edd})$, with  $r=0.69~(P=10^{- 10})$ and $\tau=0.52~(P < 10^{- 26})$.
Running the correlation analysis separately for the NLS1s and BLS1s in the clean sample, positive trends with different slopes are found at higher significance levels compared to those found for the NLS1s and BLS1s in the sample comprising sources with warm absorbers.  
$\Gamma_{\rm NLS1,clean} = (2.10\pm 0.01) + (0.40\pm 0.02) \log(L_{\rm bol}/L_{\rm Edd)}$, with $r=0.81~(P=5\times 10^{- 8})$ and $\tau=0.63~(P =7\times10^{- 7})$, 
and $\Gamma_{\rm BLS1,clean} = (1.77\pm 0.02) + (0.06\pm 0.01) \log(L_{\rm bol}/L_{\rm Edd})$, with $r=0.40~(P=5\times 10^{- 3})$ and $\tau=0.30~(P =3\times10^{- 3})$.
This correlation is illustrated in Figure~\ref{figure:fig5}.

\begin{figure*}
	\includegraphics[width=0.69\columnwidth]{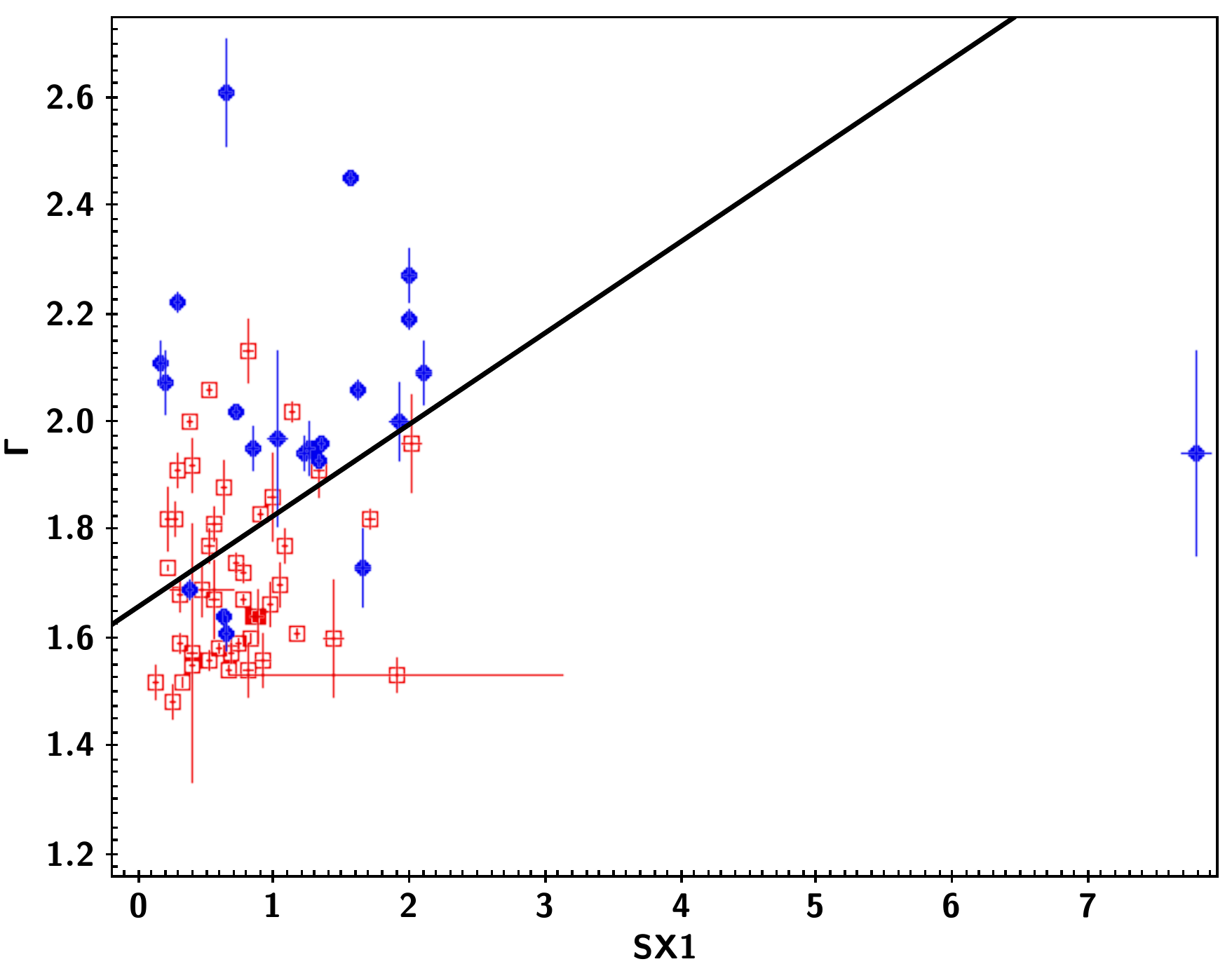}
	\hfill
	\includegraphics[width=0.69\columnwidth]{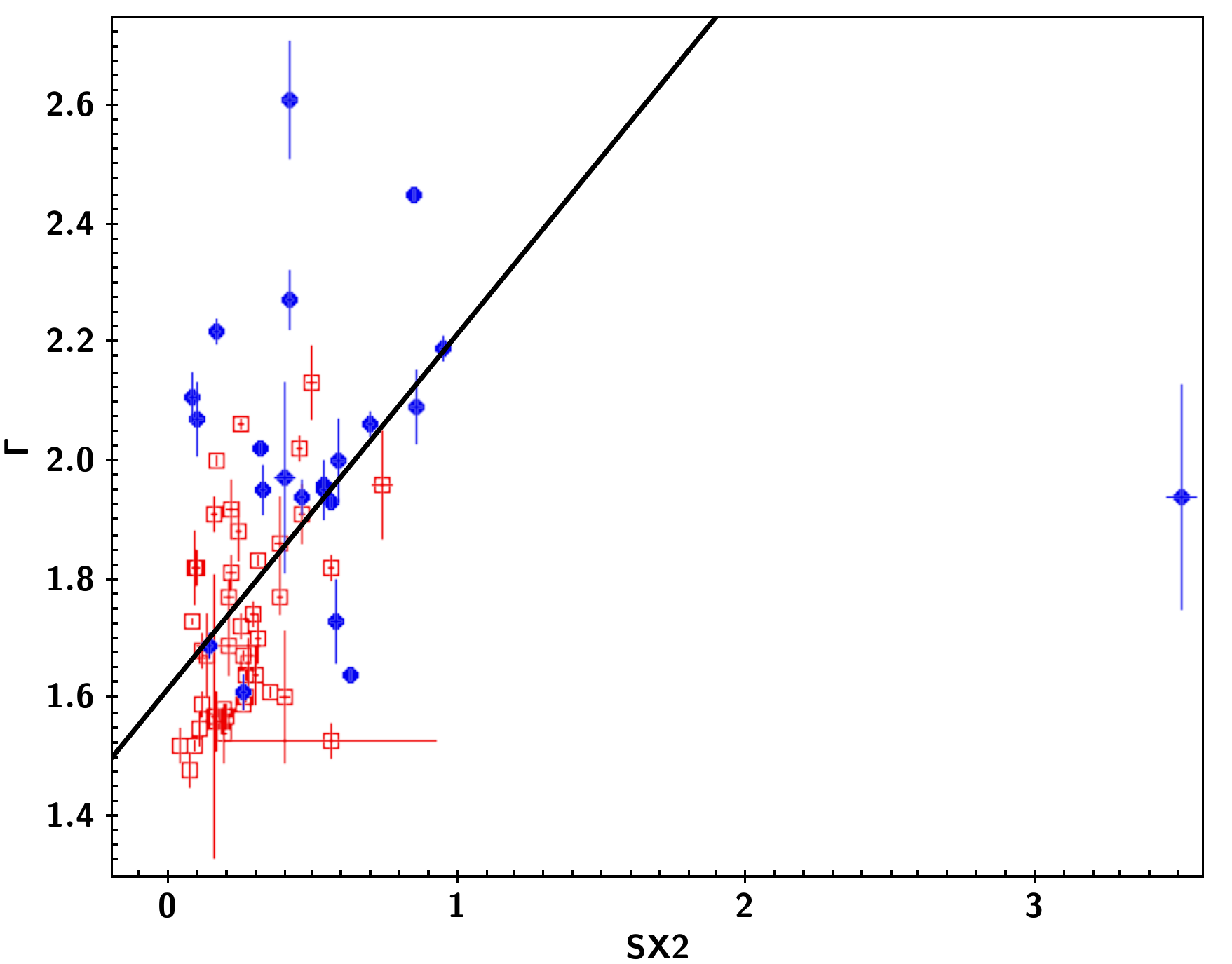}
	\hfill
	\includegraphics[width=0.69\columnwidth]{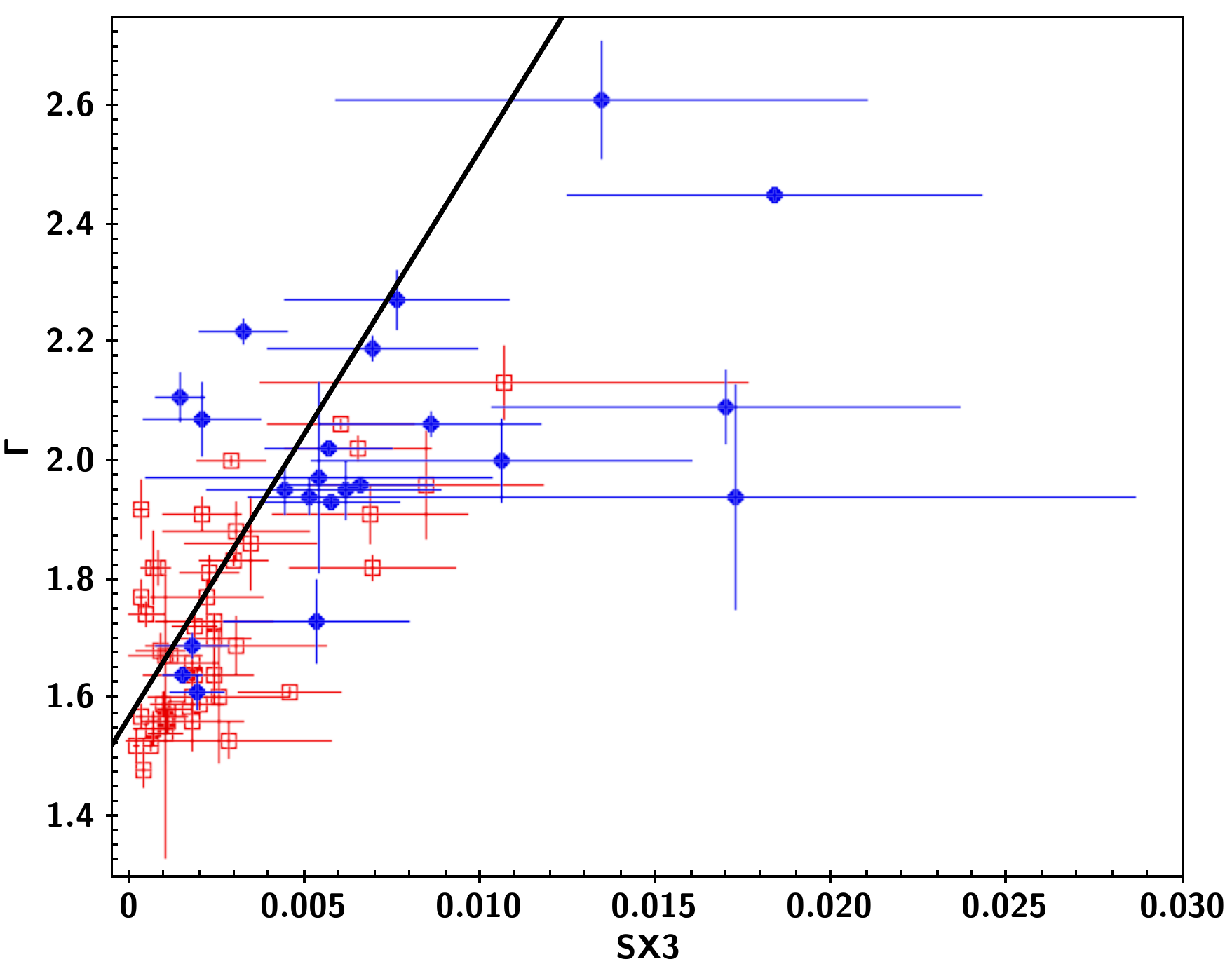}
	\caption{$\Gamma$ plotted vs. SX1, SX2, and SX3. The thick continuous line represents the linear regression fit obtained excluding the outlier RE~J1034+396 (represented by the isolated point  on the far right of the left and middle plots, and by the point with the largest x uncertainty on the right plot). The same conclusions with a slightly flatter slope are obtained when the outlier is included in the analysis.}
	\label{figure:fig6}
\end{figure*}

\subsection{Photon index vs. soft excess strength}
Figure~\ref{figure:fig6} shows the photon index $\Gamma$  vs. the soft excess strength measurements ($SX1$ in the left panel, $SX2$ in the middle, and $SX3$ in the right panel). All three plots reveal the presence of a positive correlation, whose statistical significance increases from $SX1$ to $SX3$. Specifically, we find $\Gamma = (1.66\pm 0.01) + (0.17\pm 0.01) SX1$, with  $r=0.26~(P=3\times 10^{- 2})$ and $\tau=0.19~(P =2.2\times 10^{- 2})$; $\Gamma = (1.62\pm 0.01) + (0.60\pm 0.02) SX2$, with  $r=0.45~(P=10^{- 4})$ and $\tau=0.33~(P =6\times 10^{- 5})$; and $\Gamma = (1.57\pm 0.01) + (97.4\pm 2.4) SX3$, with  $r=0.72~(P= 10^{- 11})$ and $\tau=0.54~(P < 10^{- 26})$. When we carry out the same analysis on the individual samples of NLS1s and BLS1s, the positive correlation remains statistically significant for $SX2$ in the BLS1s and for $SX3$ in both NLS1s and BLS1s. On the other hand, $SX1$ in NLS1s and BLS1s separately, and $SX2$ in NLS1s, do not show any significant correlation with $\Gamma$.
To be conservative, we have  excluded the outlier RE~J1034+396 (shown on the far right of the left and middle plots) when we computed the linear regression fits quoted above. Including the outlier does not change the results substantially; its only effect is to slightly decrease the slope value and slightly increase the significance of the correlation.

\subsection{Soft excess strength vs. accretion rate}
\begin{figure*}
	\includegraphics[width=0.69\columnwidth]{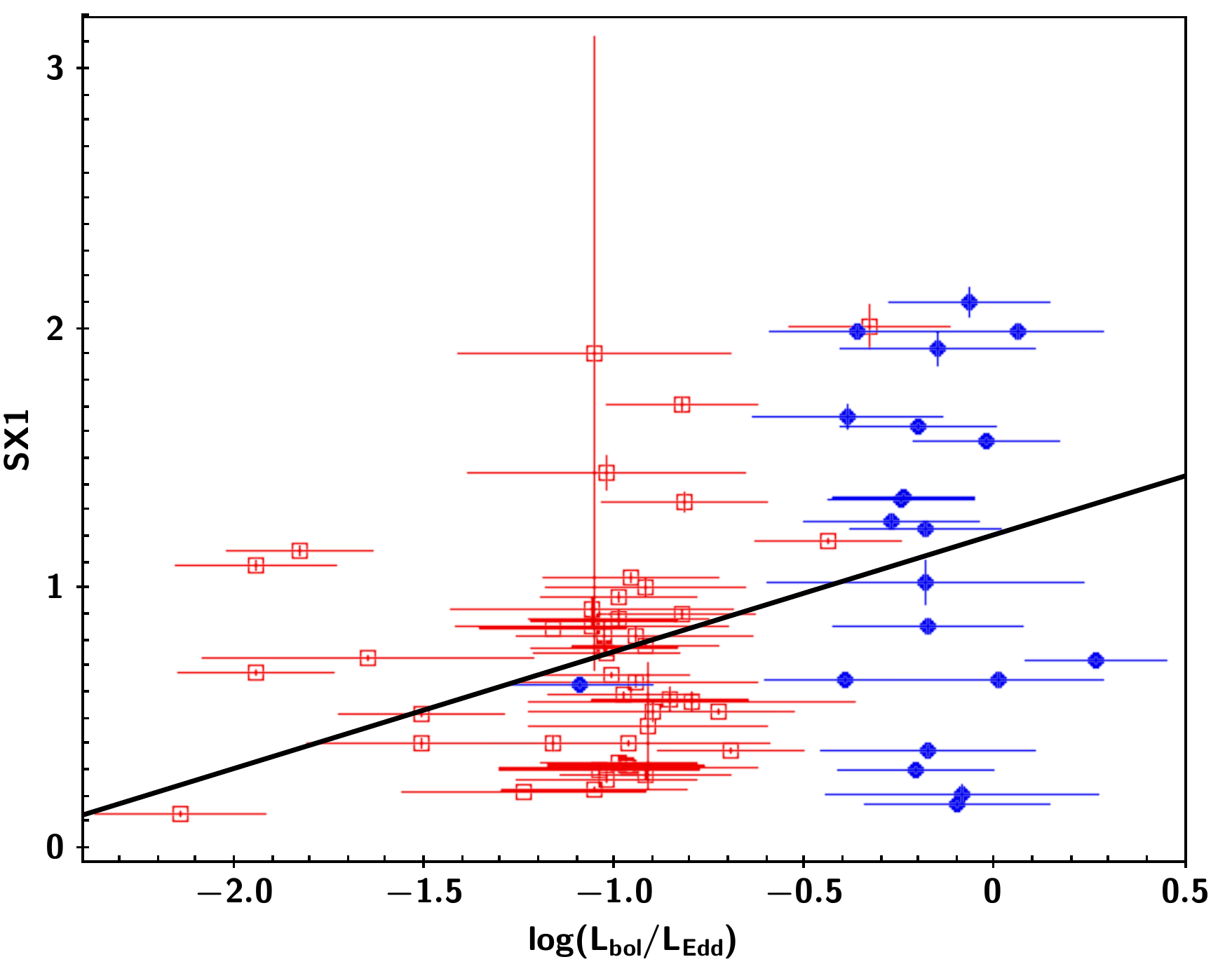}
	\hfill
	\includegraphics[width=0.69\columnwidth]{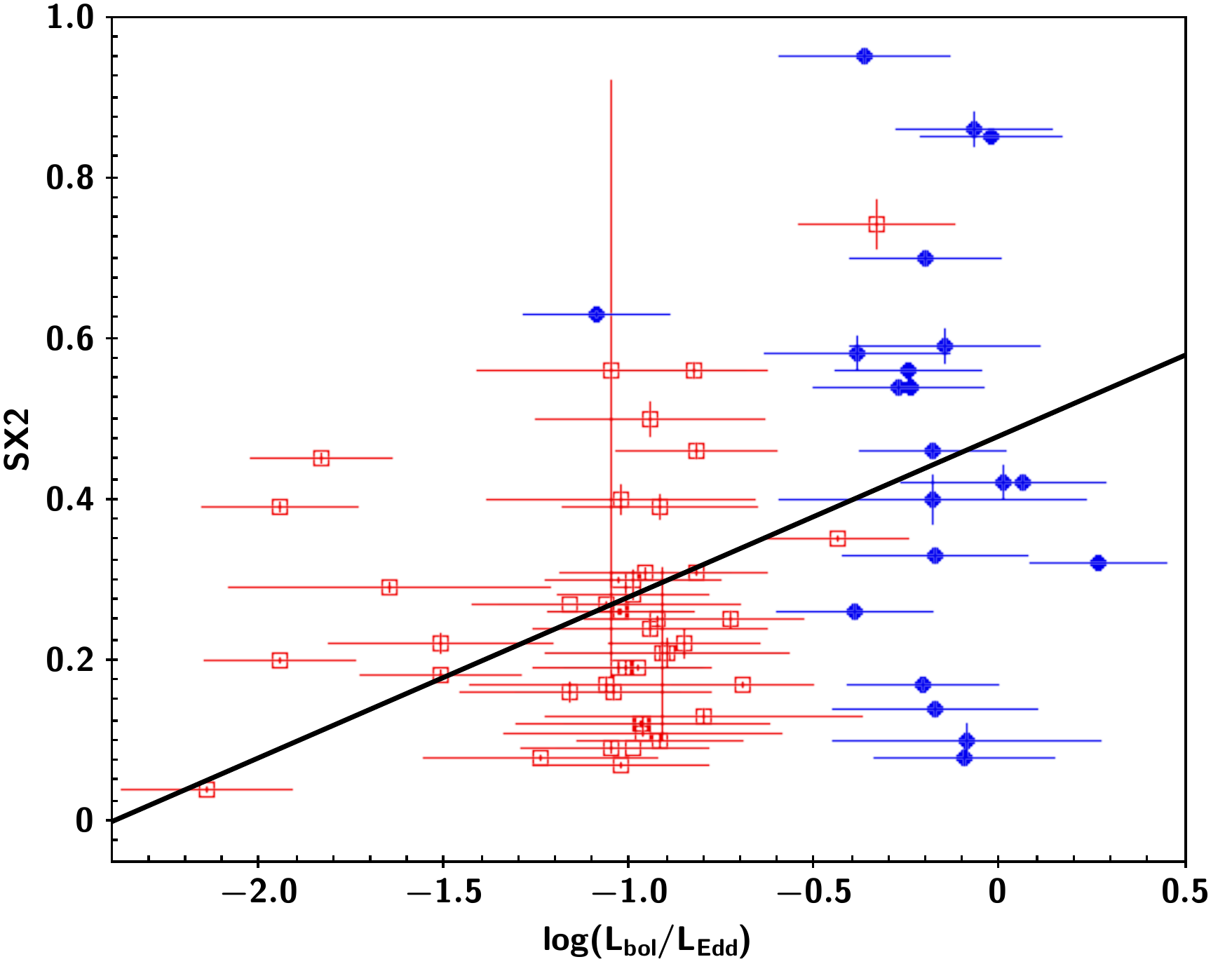}
	\hfill
	\includegraphics[width=0.69\columnwidth]{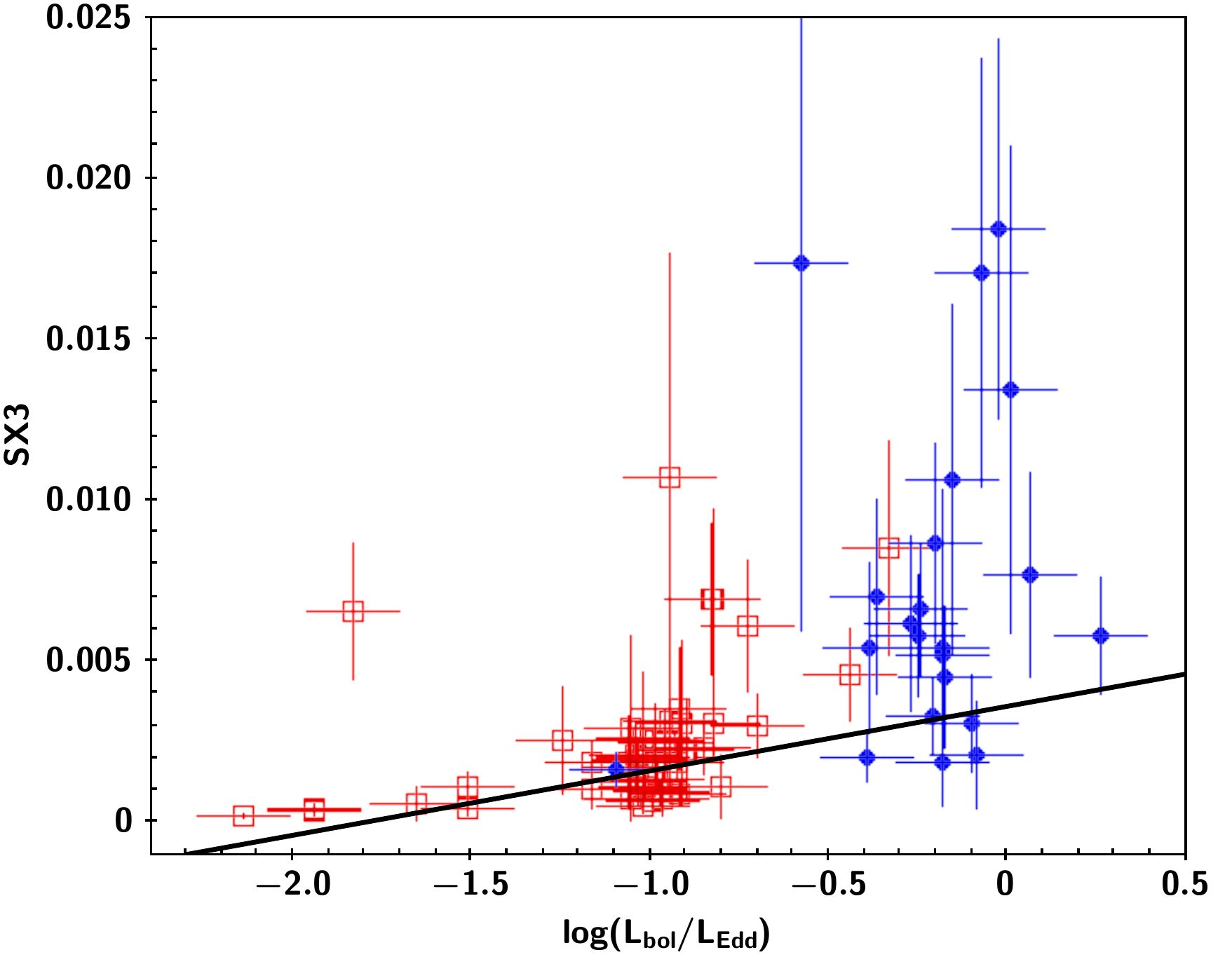}
	\caption{Soft excess strengths plotted vs. $\log(L_{\rm bol}/L_{\rm Edd})$. The thick continuous line represents the linear regression fit obtained excluding the outlier RE~J1034+396 (not represented in these plots). The same conclusions with a slightly steeper slope are obtained when the outlier is included in the analysis.}
	\label{figure:fig7}
	\end{figure*}
Figure~\ref{figure:fig7} shows the soft excess strength measurements ($SX1$ in the left panel, $SX2$ in the middle, and $SX3$ in the right panel) plotted versus $\log(L_{\rm bol}/L_{\rm Edd})$. All three plots reveal the presence of a positive correlation (with substantial scatter) whose statistical significance increases from marginally significant for $SX1$ to very significant for $SX3$. 
Specifically, we find $SX1 = (1.21\pm 0.02) + (0.45\pm 0.02) \log(L_{\rm bol}/L_{\rm Edd)}$, with  $r=0.29~(P=1.9\times 10^{- 2})$ and $\tau=0.20~(P =1.5\times 10^{- 2})$; $SX2 = (0.48\pm 0.01) + (0.20\pm 0.01) \log(L_{\rm bol}/L_{\rm Edd)}$, with  $r=0.36~(P=2.5\times 10^{- 3})$ and $\tau=0.25~(P =2.7\times 10^{- 3})$; and $SX3 = (0.0036\pm 0.0001) + (0.0020\pm 0.0005) \log(L_{\rm bol}/L_{\rm Edd)}$, with  $r=0.68~(P= 10^{- 10})$ and $\tau=0.50~(P < 10^{- 26})$.
 When we carry out the same analysis on the individual samples of NLS1s and BLS1s, there is no significant correlation with the exception of $SX3$, which remains positively correlated with $\log(L_{\rm bol}/L_{\rm Edd)}$ at high significance level only for the BLS1s sample.
To be conservative, we have  excluded the outlier RE~J1034+396 (which for clarity is not shown in the plots) when we computed the linear regression fits quoted above. Including the outlier does not change the results substantially and its only effect is to slightly increase the slope value and slightly increase the significance of the correlation.

\subsection{Soft excess strength vs. black hole mass}
\begin{figure*}
	\includegraphics[width=0.69\columnwidth]{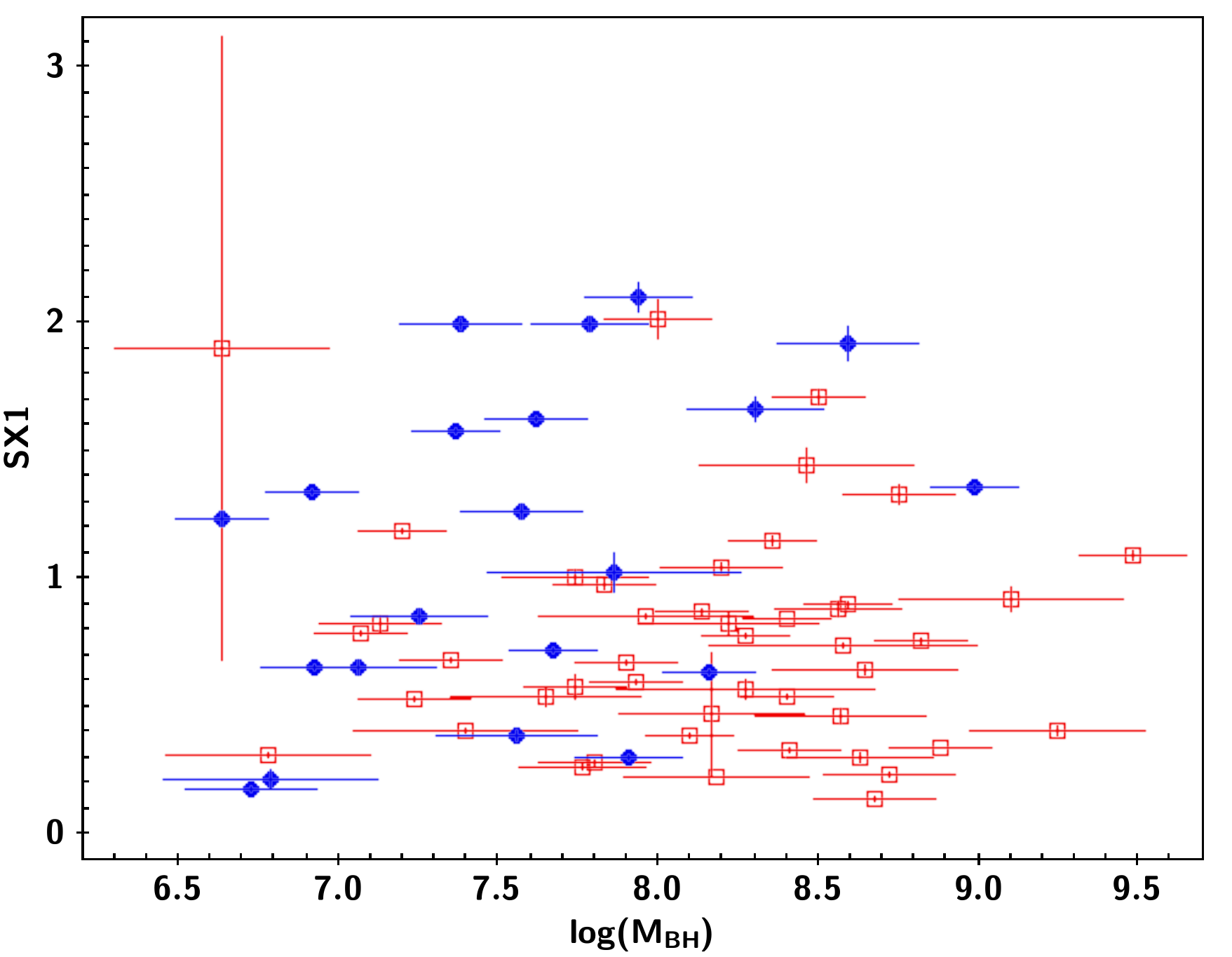}
	\hfill
	\includegraphics[width=0.69\columnwidth]{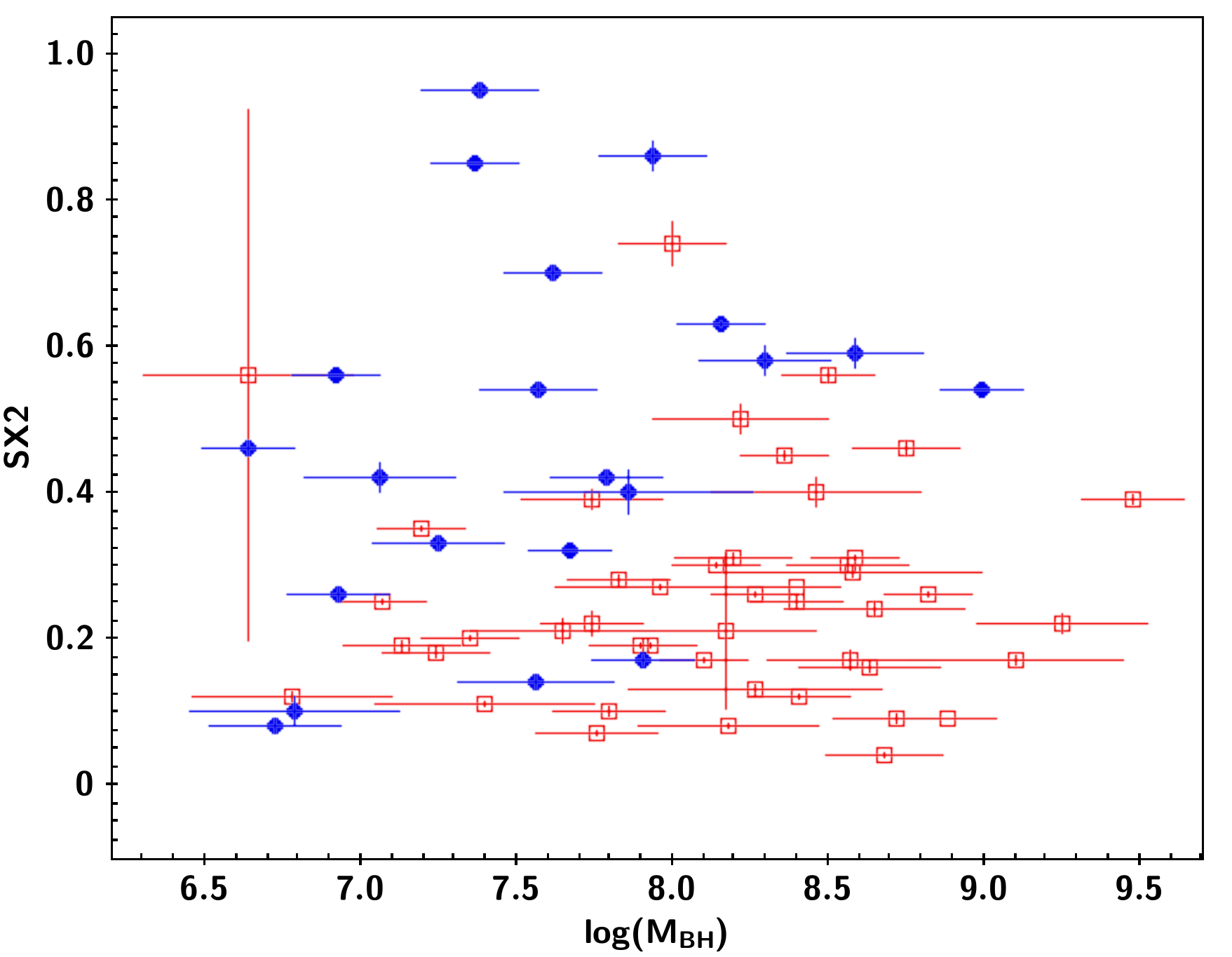}
	\hfill
	\includegraphics[width=0.69\columnwidth]{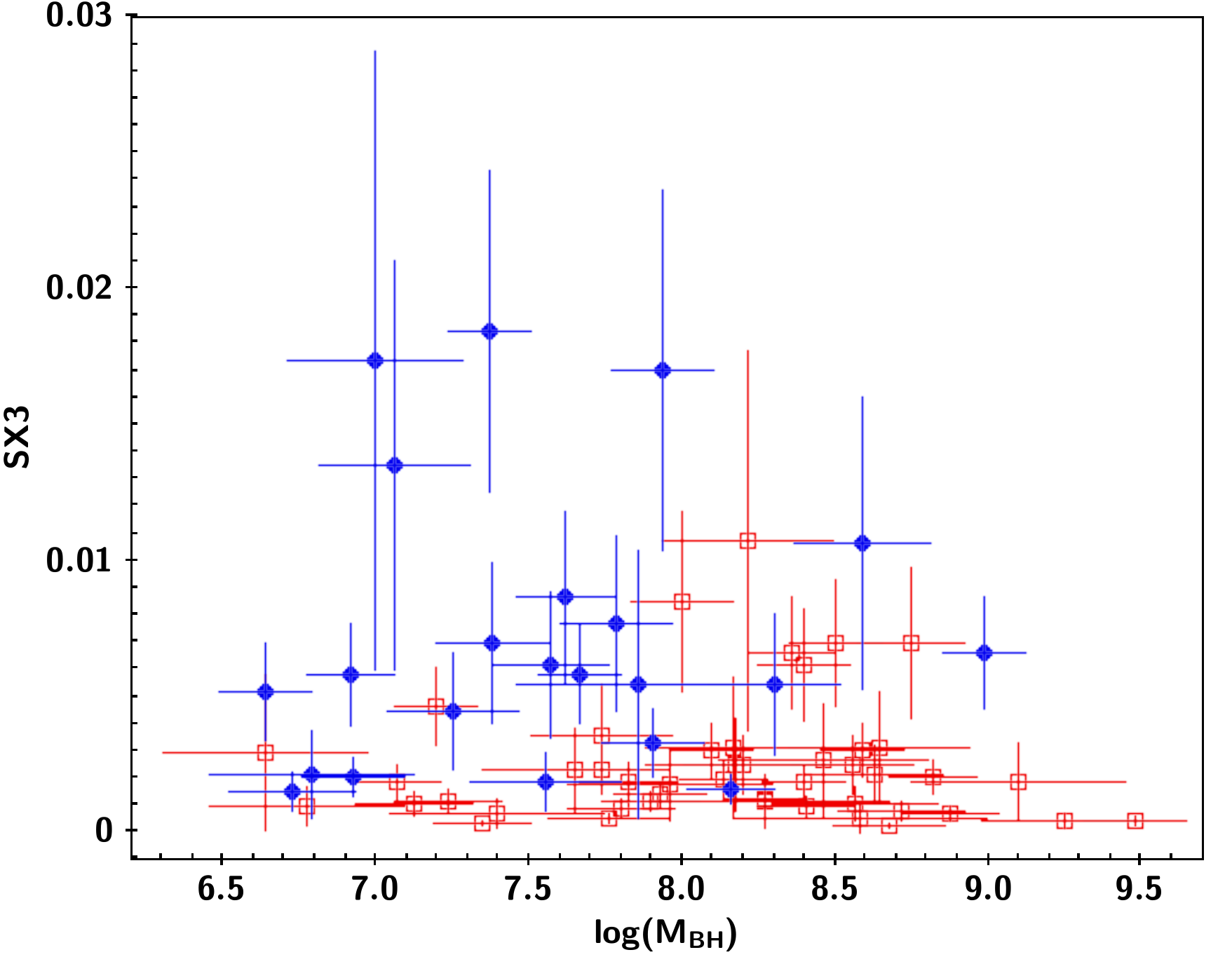}
	\caption{Soft excess strengths plotted vs. $\log(M_{\rm BH})$. No clear correlation exists in any of the plots. For the sake of clarity, the outlier RE~J1034+396 was not represented in these plots.}
	\label{figure:fig8}
\end{figure*}
Figure~\ref{figure:fig8} shows the soft excess strength measurements ($SX1$ in the left panel, $SX2$ in the middle, and $SX3$ in the right panel) plotted versus $\log(M_{\rm BH})$. A visual inspection of these plots does not reveal any clear trend. Indeed, the linear regression analysis suggests the presence of weak negative trends in all cases, and Spearman's and Kendall's rank correlation analyses confirm that there is no statistically significant correlation. Specifically, for $SX1$ $r= -0.05~(P=0.69)$ and $\tau=-0.02~(P =0.73)$; for $SX2$ $r= -0.09~(P=0.47)$ and $\tau= -0.07~(P =0.43)$; and for $SX3$ $r= -0.16~(P=0.18)$ and $\tau= -0.11~(P =0.15)$.

\subsection{Soft excess strength vs. X-ray luminosity}
\begin{figure}
	\includegraphics[width=0.99\columnwidth]{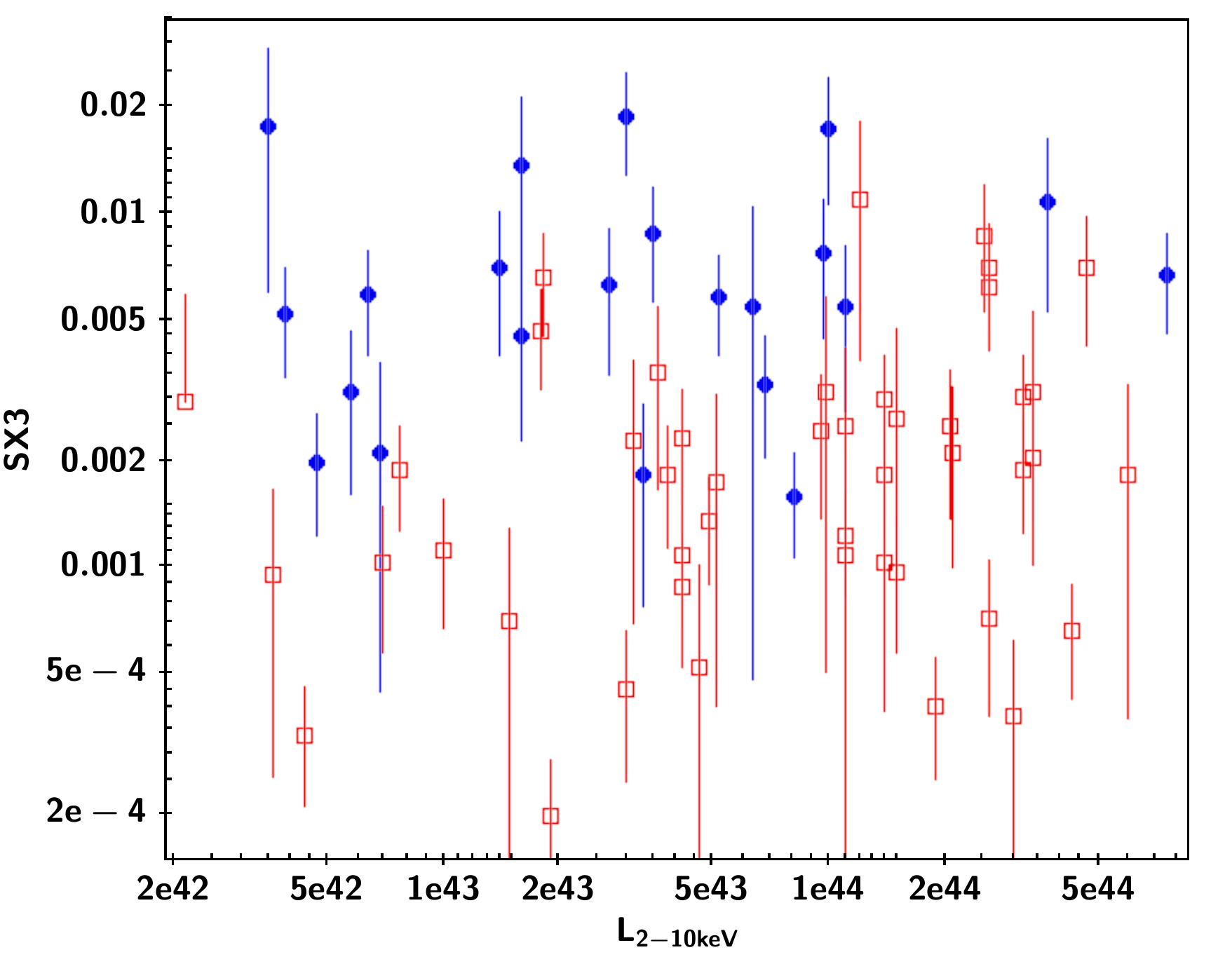}	
	\caption{Soft excess strength plotted vs. $L_{\rm 2-10~keV}$. No clear correlation exists bewtween $SX3$ and the 2-10 keV luminosity of the primary component.}
	\label{figure:fig9}
\end{figure}
Important information about the nature of the soft excess can be obtained by studying the correlation between the soft excess strength and the luminosity associated with the primary emission. Generally, because the definition of the soft excess explicitly contains the hard X-ray luminosity (or flux) in the denominator, such  analysis cannot be performed directly. However, the soft excess strength parametrized by $SX3$, which is fully consistent with $SX1$ and $SX2$ (see Section 4) does not explicitly contain the primary X-ray emission, allowing for a direct comparison.
Figure~\ref{figure:fig9} shows the soft excess strength $SX3$ plotted versus the 2--10 keV luminosity of the primary continuum. A visual inspection of this figure reveals the absence of any trend, in addition to showing that the NLS1s and BLS1s in our sample span the same luminosity range and are clearly separated in soft excess strength. Indeed, Spearman's and Kendall's rank correlation analyses confirm that there is no statistically significant correlation, with $r= 0.02~(P=0.84)$ and $\tau=0.01~(P =0.89)$.

\section{Discussion and Conclusions}
The nature of the soft X-ray excess is still very much debated. Initially, after its first detection \citep{prav81} and  after it became clear that this feature was fairly common among type 1 AGN, the soft excess was generally explained as thermal emission from the innermost part of the accretion disk \citep[e.g.,][]{turn89}. However, the discrepancy between the uniform temperatures detected in AGN samples with different BH masses and accretion rate values  and the expected dependence of the disk temperature,  $T^4 \propto \dot m /M_{\rm BH}$ (where $\dot m$ is the accretion rate in Eddington units), 
led to the conclusion that the soft excess could not be simply produced by thermal emission and to the hypothesis that atomic processes could be the cause of the soft excess. For example, \citet{gierl04} hypothesized that the soft excess could be explained by relativistically smeared absorption from disk winds. On the other hand, \citet{crum06} proposed an alternative model able to explain the soft excess in terms of relativistically blurred ionized reflection from the accretion disk. Both models have been successfully used to characterize the soft excess in numerous AGN; however, both require extreme parameters. In particular, the relativistic absorption model requires very high wind velocities, which are hard to achieve in accretion disks \citep{schur09}, whereas the relativistic reflection models tend to require maximally rotating BHs, very steep disk emissivity laws, and light-bending effects \citep[e.g.,][]{mini04} to explain the soft excess consistently with other reflection features. An alternative model to explain the soft excess is the presence of a warm Comptonizing corona, in addition to the postulated hot corona responsible for the primary X-ray emission \citep[e.g.,][]{magdz98}. In the latter case, the challenge is to explain the constancy of the soft excess temperature for a broad range of \mbh\ values.

In this work, we analyzed a sample of 89 type 1 AGN (59 BLS1s and 30 NLS1s), originally derived from a flux-limited sample observed with \xmm. For both BLS1s and NLS1s, in paper I we determined the \mbh\ in a homogeneous way using an X-ray scaling method, which yields values fully consistent with the reverberation mapping ones, but is independent of the inclination angle of the system and makes no assumption on the geometry of the broad-line region \citep{glioz11}. For all objects, we also determined the $\lambda_{\rm Edd}=(L_{\rm bol}/L_{\rm Edd})$ values, using the bolometric corrections prescribed by \citet{vasu09} for BLS1s and NLS1s. Although this is by no means a complete sample, it offers the possibility to compare physical properties in two well-defined samples of NLS1s and BLS1s that are matched in X-ray luminosities and have masses and accretion rates homogeneously and robustly constrained. In our sample, NLS1s are very clearly separated from BLS1s in terms of $\lambda_{\rm Edd}$ distributions, and NLS1s on average have smaller \mbh\ values than BLS1s despite a substantial overlap between their distributions (see Fig.~\ref{figure:fig1}).

Of the original sample of 89 AGN, we retained 68 objects (46 BLS1s and 22 NLS1s) in our final clean sample, after excluding the objects whose soft X-ray emission is severely affected by the presence of warm absorbers that hamper the characterization of the soft excess. In our starting sample the fraction of warm absorbers appears to be slightly larger in NLS1s (27\%) than in BLS1s (22\%).
The presence of soft excess was revealed in every single object of our sample with different apparent strengths (see Fig.~\ref{figure:fig2}), confirming that this feature is truly ubiquitous in type 1 AGN. To investigate the nature of the soft excess we did not fit a specific physically motivated model; instead we fitted the soft X-ray part of the spectrum with a phenomenological model represented by one or two blackbody components and then computed its flux to measure the soft excess strength. We then systematically compared the soft excess strength in the two samples, and carried out a correlation analysis to test whether statistically significant trends exist between the soft excess and the relevant parameters of the black hole systems.

To quantify the strength of the soft excess we used three different quantities. $SX1=({F_{\mathrm{bb}}}/{F_{\mathrm{bmc}}})_{0.5-2\ \mathrm{keV}}$, which compares the flux of the blackbody model to that of the Comptonization model extrapolated in the 0.5--2 keV band; this parametrization has been frequently used in the literature and allows a direct comparison with previous works, but it is sensitive to the normalization of the Comptonization model in the soft band, which may not be accurate, considering that the Comptonization model parameters have been obtained using only the 2--10 keV spectrum and then some parameters have been frozen before the extrapolation to the soft band. Our second parametrization of the soft excess, $SX2=({F_{\mathrm{bb}}}/{F_{\mathrm{bmc}}})_{0.5-10\ \mathrm{keV}}$, is similar to $SX1$ but more robust, because it compares the strengths of the blackbody and Comptonization components in the whole 0.5--10 keV range. Finally, we used $SX3=(L_{\mathrm{bb}_{0.5-2\ \mathrm{keV}}}/L_{\mathrm{Edd}})$, which we regard as the least model-dependent of the three measurements, since the Comptonization component does not enter into its definition. These three different parametrizations of the soft excess strength appear to be fully consistent with each other, as indicated by Fig.~\ref{figure:fig3}, which shows that $SX1$, $SX2$, and $SX3$ are tightly correlated; this is formally confirmed at high significance level by Spearman's and Kendall's rank correlation analyses.

The systematic comparison of the soft excess strength in NLS1s and BLS1s is illustrated in Fig.~\ref{figure:fig4}, which shows histograms of $SX1$, $SX2$, and $SX3$ and suggests that the distributions of soft excess strength are different in NLS1s and BLS1s regardless of the parametrization used. Fig.~\ref{figure:fig4} also reveals the presence of one outlier in the $SX1$ and $SX2$ distributions of NLS1s, with soft excess strength at least three times larger than any other value. This outlier, RE J1034+396, which is one of the very few AGN in which quasi-periodic oscillations (QPOs) have been robustly detected, was already known for its extreme soft excess \citep{midd09}. To be conservative, we have carried out the statistical comparison between  NLS1s and BLS1s excluding RE J1034+396. In any case, the inclusion of the outlier in the analysis does not affect the general conclusions. Based on K-S and Student's $t$ tests, the soft excess strength in NLS1s is significantly higher than in BLS1s, although there is substantial overlap between their distributions. A similar but qualitative conclusion was reached by \citet{midd07} using a considerably smaller sample of AGN and with a different parametrization of the soft excess strength.  The same conclusion -- narrow-line type 1 AGN have stronger soft excess than broad-line type 1 AGN -- at high significance level was also obtained by \citet{bian09a} based on the CAIXA catalog, which contains  77 quasars (16 narrow-line and 35 broad-line objects) and 79 Seyfert galaxies (21 narrow-line and 30 broad-line sources).

The existence of statistically significant correlations between physical quantities and relevant parameters of black hole systems plays an important role in our understanding of BH systems and their central engines. For example, a positive correlation between the photon index $\Gamma$ and the accretion rate in Eddington units 
$\lambda_{\rm Edd}$ has been robustly determined in different classes of AGN accreting at moderate or high level (e.g., \citealt{shem08,risa09,bright13,bright16}, but see \citealt{trak17} for a discording view). This result, which is strengthened by the steeper-when-brighter trend consistently observed in long-term spectral variability studies of AGN \citep[e.g.,][]{sobo09} and in large samples of quasars \citep{sera17}, suggests that the photon index can be used as an indicator of the accretion rate. This is consistent with the behavior regularly observed in Galactic black holes (GBHs) during their spectral transitions between the low/hard and high/soft states \citep[e.g.,][]{remi06}, and lends support to the unification model of black hole systems at all scales, where the different classes of AGN can be associated with different spectral states of Galactic black holes (GBHs) during their spectral evolution \citep[e.g.,][]{done05}.

Using our whole sample of 89 AGN (note that the same conclusions are obtained using the clean sample, as shown in Section 5.1), we have performed a correlation analysis of $\Gamma$ vs. $\log(\lambda_{\rm Edd})$ and found a positive linear correlation at high significance level, described by the equation $\Gamma = (2.09\pm 0.05) + (0.35\pm 0.05) \log(\lambda_{\rm Edd})$, which is broadly consistent with the results obtained by \citet{shem08}, \citet{risa09}, and \citet{bright13}. When we carried out the analysis on NLS1s and BLS1s separately, we still found positive highly significant correlations with a steeper slope for NLS1s compared to BLS1s. A similar conclusion, at lower significance, was obtained by \citet{wang04} using a sample of 56 type 1 AGN observed with ASCA.
As discussed in Section 3, a Compton reflection component is not included in our spectral analysis because the \xmm\ energy range hampers the characterization of such a component. We note that \citet{bian09a} explicitly address this issue by including a luminosity-dependent Compton reflection component, which has the effect of slightly increasing the values of the photon index for both narrow- and broad-line objects. Interestingly, our average values of $\Gamma$, reported in Section 2, are fully consistent with the values quoted by \citet{bian09a} after they added the reflection component, suggesting that the effect of the reflection component is negligible in our sample.

Using our clean sample of 68 AGN, we investigated the existence of trends between the soft excess strengths, as measured by $SX1$, $SX2$, and $SX3$, and different parameters that characterize the black hole systems. When the photon index is plotted vs. the soft excess strength a positive trend is observed in all plots (see Fig.~\ref{figure:fig6}). This positive correlation is marginally significant for $SX1$ (at significance level of $\sim$97\%) and very significant for $SX2$, and $SX3$ (significance > 99.9 \%). Since we regard $SX3$ as the most robust measurement of soft excess strength, we conclude that there is a positive correlation between $\Gamma$ and the soft excess strength. Our results are in agreement with the strong, positive correlation found by \citet{bian09b} using the CAIXA catalog.
Similar results were also found by \citet{bois16} using a sample of 102 Seyfert 1 galaxies from the Swift BAT 70-month catalog. Since these authors use the same definition as our $SX1$ to parametrize the soft excess strength (although the phenomenological model they used to fit the soft X-rays is Bremsstrahlung, as opposed to the blackbody model utilized in this work), we can directly compare their linear regression results (shown in their Eq. 4) and find that it is fully consistent with our best linear fit reported in section 5.2. Interestingly, \citet{bois16}  also show that the correlation predicted by the ionized reflection model yields a weak negative slope. Therefore, our results -- the statistically significant positive correlation between $\Gamma$ and the soft excess strength -- confirm and strengthen their conclusion that the soft excess in the bulk of type 1 AGN is unlikely to be produced by relativistically blurred ionized reflection. This does not rule out that ionized reflection may be the physical explanation for the soft excess in objects whose X-ray emission is affected by strong gravitational bending, such as MCG-6-30-15 \citep{chia11}; indeed, this conclusion appears to be independently confirmed by X-ray time lag analysis \citep
{emma11}.

In our work, we also found another positive trend when the soft excess strength ($SX1$, $SX2$, and $SX3$) is plotted versus the Eddington ratio (see Fig.~\ref{figure:fig7}). This is expected, given the strong correlation found between $\Gamma$ and the Eddington ratio $\lambda_{\rm Edd}$, and the positive trends observed between $\Gamma$ and the soft excess strength. Similarly to the previous analysis, the linear correlation is highly significant for $SX3$, significant at 99\% confidence level for $SX2$, and marginally significant ($\sim$98\%) for $SX1$. 
A similar conclusion was also reached by \citet{bois16} but not by 
\citet{bian09b}. The apparent discrepancy with the latter can be explained by the fact that \citet{bian09b} do not exclude from their analysis sources with warm absorbers, as opposed to our approach and that of \citet{bois16}. Additionally, the CAIXA sample only has black hole mass values for 57\% of the sample and they are obtained from different indirect methods, whereas we have calculated the black hole masses for all objects of our sample in a homogenous way with an X-ray scaling method, which provides \mbh\ values in agreement with the reverberation mapping method. Finally, \citet{bian09b} use a luminosity-dependent bolometric correction as opposed to the accretion rate-dependent one used in our work. All these differences (in addition to the different baseline spectral model) can explain the different conclusion reached by \citet{bian09b}. In addition, we note that the positive correlation between soft excess strength and accretion rate is only marginally significant when we use $SX1$, which is close to the parametrization of the soft excess strength used by \citet{bian09b}. 

No evident trend nor correlation is found when the soft excess strength is plotted versus the black hole mass (see Fig.~\ref{figure:fig8}), which is in agreement with previous systematic analysis of soft excess in AGN samples \citep[e.g.,][]{pico05,bian09b}.

Finally, no positive or negative correlation is found when the soft excess strength measured by $SX3$ is plotted versus the 2--10 keV luminosity of the primary component (see  Fig.~\ref{figure:fig9}). This is an important result because it rules out in a model-independent way that the soft excess is primarily produced by reflection or absorption. Indeed, both scenarios assume that the primary X-ray emission is strongly suppressed either by strong light bending or by strong absorption from outflowing winds. As a consequence, as pointed out by \citet{bian09b}, both scenarios predict an anticorrelation between the soft excess strength and the primary X-ray emission, parametrized by the 2--10 keV luminosity. Since our $SX3$ parameter does not include directly $L_{\rm 2-10~keV}$, we are able  to test this correlation directly for the first time. The clear lack of an inverse correlation, formally confirmed by Spearman's and Kendall's rank correlation analyses, strongly argues against the reflection and absorption scenarios.

Overall, the correlations (or lack thereof) derived in this work appear to be at odds with those predicted by the ionized reflection model and are instead naturally explained in the framework of the warm Comptonization model, where some of the optical/UV seed photons thermally produced by the accretion disk are Comptonized by a warm optically thick region, which may be an upper layer of the accretion disk itself \citep[e.g.,][]{petru18} or an inner converging flow \citep[e.g.,][]{tita97}, before being Comptonized in an optically thin hot corona. In this scenario, an increase of the accretion rate would lead to an enhancement in the seed photon flux, which in turn would produce a stronger soft excess via the warm Comptonization component. Part of the soft excess flux would be up-scattered by the hot electrons in the corona to produce the primary X-ray continuum and eventually cool the corona, which would explain the correlation between the photon index and the soft excess strength.
In this framework, the lack of variation of the soft excess strength with \mbh\ may be explained by bulk motion Comptonization, or by thermal Comptonization, provided that the accretion power released in the disk and in the corona vary in concert in such a way to produce the proper physical conditions with optical depth $\tau > 1$ and $kT\sim 1$ keV \citep[e.g.,][]{done12,roza15,petru18}. This conclusion is supported and strengthened by recent findings obtained applying physically motivated warm Comptonization models to fit high-quality broad-band spectra of different AGN, obtained from simultaneous long observations carried out by \xmm\ and {\it NuSTAR} 
\citep[e.g.,][]{porq18,midde18,urs18}

We conclude by summarizing the main findings of our work. Starting from a flux-limited sample of 89 type 1 AGN (59 BLS1s and 30 NLS1s), with BLS1s and NLS1s matched in X-ray luminosities, and with NLS1s with lower \mbh\ on average and considerably larger $\lambda_{\rm Edd}$ values than BLS1s, we defined a clean sample of 68 objects (46 BLS1s and 22 NLS1s) after excluding 13 BLS1s and 8 NLS1s severely affected by warm absorbers that hamper the proper characterization of the soft excess. Using the clean sample we obtained the following results:
\begin{itemize}
\item The soft excess is ubiquitously detected in both BLS1s and NLS1s with a 100\% detection rate.
\item  The strength of the soft excess (which has been parametrized in three different ways) is significantly larger in the NLS1 sample, compared to the BLS1 sample, regardless of the parametrization used.
\item  Combining BLS1s and NLS1s, the strength of the soft excess is positively correlated with the photon index $\Gamma$.
\item  Similarly, the strength of the soft excess appears to positively correlate with the accretion rate.
\item  Conversely, there is no correlation at all between the strength of the soft excess and black hole mass.
\item  Importantly, no inverse correlation between the soft excess strength and the X-ray primary continuum was found.
\end{itemize}

The results from the correlation analysis appear to favor the warm Comptonization scenario as the origin of the soft excess. However, larger, more complete samples  are needed to confirm these conclusions.

\section*{Acknowledgements}
We thank the anonymous referee for constructive comments and suggestions that improved the clarity of the paper and helped strenghten our conclusions.








\appendix

\section{Additional Spectral Results}

There were 21 AGN (13 BLS1s and 8 NLS1s) that required warm absorbers (which we parametrized with one or two \texttt{zxipcf} models in \xspec) to get an acceptable fit. As we explained in Section 3, that precludes the measurement of the soft excess strength, and that is why those 21 AGN do not appear in the soft excess strength tables. 

The eight NLS1s that required warm absorbers were Mrk 335 \citep{gall13}, NGC 4051 \citep{mizu17, nuci10}, Mrk 766 \citep{buis18, emma11}, Was 61 \citep{dou16}, MCG-6-30-15 \citep{chia11, emma11, kamm17}, NGC 5506 \citep{sun18}, PG 1448+273, and IRAS 17020+4544 \citep{long15, leig97, komo98}. 

The 13 BLS1s that required warm absorbers were Ark 120 \citep{nard11, porq18}, H 0557-385 \citep{long09}, PG 0844+349 \citep{poun04}, NGC 3516 \citep{huer14, cost11, mehdi10}, PG 1114+445 \citep{asht04}, NGC 3783 \citep{blus02}, Ark 374, IC 4329A \citep{stee05}, NGC 5548 \citep{mao18}, Mrk 1383, Mrk 876 \citep{porq04}, Mrk 304 \citep{pico04, brin04}, and Fairall 1146.

Tables~\ref{tab:3} and \ref{tab:4} show the spectral results for NLS1s and BLS1s obtained using our baseline model.
In Tables~\ref{tab:5} and \ref{tab:6} we report the soft excess strength measurements for NLS1s and BLS1s, respectively.

\begin{table*}
	\caption{NLS1 spectral data}
	\begin{center}
		\begin{tabular}{lrrrr} 
			\toprule
			\toprule       
			\multicolumn{1}{c}{Name} 
			& \multicolumn{1}{c}{$kT_\mathrm{BB,1}$} & \multicolumn{1}{c}{$kT_\mathrm{BB,2}$} & \multicolumn{1}{c}{$\Gamma$} & \multicolumn{1}{c}{$\chi^2$/d.o.f.} \\
			\multicolumn{1}{c}{(1)} 
			& \multicolumn{1}{c}{(2)} & \multicolumn{1}{c}{(3)}
			& \multicolumn{1}{c}{(4)} & \multicolumn{1}{c}{(5)} \\
			\midrule
			I Zw 1 & $0.08_{-0.01}^{+0.01}$ & $\dots$ & $2.22\pm0.02$ & $959/880$ \\
			\noalign{\smallskip}
			Ton S180 & $0.16_{-0.01}^{+0.01}$ & $0.08_{-0.01}^{+0.01}$ & $2.06\pm0.02$ & 1250/1199 \\
			\noalign{\smallskip}
			Mrk 359 & $0.10_{-0.01}^{+0.01}$ & $\dots$ & $1.61\pm0.03$ & 734/721 \\
			\noalign{\smallskip}
			Mrk 1014 & $0.14_{-0.01}^{+0.01}$ & $\dots$ & $1.97\pm0.16$ & 166/220 \\
			\noalign{\smallskip}
			Mrk 586 & $0.09_{-0.01}^{+0.01}$ & $0.17_{-0.02}^{+0.02}$ & $2.09\pm0.06$ & 482/441 \\
			\noalign{\smallskip}
			Mrk 1044 & $0.07_{-0.01}^{+0.01}$ & $0.15_{-0.01}^{+0.01}$ & $1.93\pm0.01$ & 1506/1441 \\
			\noalign{\smallskip}
			RBS 416 & $0.08_{-0.01}^{+0.01}$ & $0.20_{-0.03}^{+0.03}$ & $1.95\pm0.04$ & 440/463 \\
			\noalign{\smallskip}
			HE 0450-2958 & $0.06_{-0.01}^{+0.01}$ & $0.19_{-0.01}^{+0.01}$ & $2.00\pm0.07$ & 401/389 \\
			\noalign{\smallskip}
			PKS 0558-504 & $0.16_{-0.01}^{+0.01}$ & $0.08_{-0.01}^{+0.01}$ & $1.96\pm0.01$ & 1611/1596 \\
			\noalign{\smallskip}
			Mrk 110 & $0.13_{-0.01}^{+0.01}$ & $0.06_{-0.01}^{+0.01}$ & $1.64\pm0.01$ & 1747/1597 \\
			\noalign{\smallskip}
			RE J1034+396 & $0.07_{-0.01}^{+0.01}$ & $0.15_{-0.01}^{+0.01}$ & $1.94\pm0.19$ & 594/630 \\
			\noalign{\smallskip}
			PG 1211+143 & $0.09_{-0.01}^{+0.01}$ & $0.19_{-0.01}^{+0.01}$ & $2.02\pm0.01$ & 1173/1157 \\
			\noalign{\smallskip}
			PG 1244+026 & $0.14_{-0.01}^{+0.01}$ & $0.05_{-0.01}^{+0.01}$ & $2.19\pm0.02$ & 994/917 \\
			\noalign{\smallskip}
			IRAS 13349+2438 & $0.08_{-0.01}^{+0.01}$ & $0.26_{-0.01}^{+0.01}$ & $2.27\pm0.05$ & 1035/891 \\
			\noalign{\smallskip}
			PG 1402+261 & $0.06_{-0.01}^{+0.01}$ & $0.16_{-0.01}^{+0.01}$ & $1.73\pm0.07$ & 481/512 \\
			\noalign{\smallskip}
			PG 1440+356 & $0.07_{-0.01}^{+0.01}$ & $0.15_{-0.01}^{+0.01}$ & $1.95\pm0.05$ & 484/526 \\
			\noalign{\smallskip}
			Mrk 493 & $0.08_{-0.01}^{+0.01}$ & $0.17_{-0.01}^{+0.01}$ & $1.94\pm0.03$ & 763/721 \\
			\noalign{\smallskip}
			Mrk 896 & $0.08_{-0.01}^{+0.01}$ & $\dots$ & $2.11\pm0.04$ & 435/417 \\
			\noalign{\smallskip}
			Mrk 1513 & $0.08_{-0.01}^{+0.01}$ & $\dots$ & $1.69\pm0.02$ & 760/708 \\
			\noalign{\smallskip}
			II Zw 177 & $0.11_{-0.01}^{+0.01}$ & $\dots$ & $2.61\pm0.10$ & 290/281 \\
			\noalign{\smallskip}
			Ark 564 & $0.07_{-0.01}^{+0.01}$ & $0.14_{-0.01}^{+0.01}$ & $2.45\pm0.01$ & 1784/1428 \\
			\noalign{\smallskip}
			AM 2354-304 & $0.10_{-0.01}^{+0.01}$ & $\dots$ & $2.07\pm0.06$ & 374/378 \\
			\bottomrule
		\end{tabular}
	\end{center}
	\begin{flushleft}
		Columns: 1 = AGN name. 2 = temperature of first blackbody in keV. 3 = temperature of second blackbody, if needed, in keV. 4 = photon index. 5 = $\chi^2$/degrees of freedom.
	\end{flushleft}
	\label{tab:3}
\end{table*}

\begin{table*}
	\caption{BLS1 spectral data}
	\begin{center}
		\begin{tabular}{lrrrr} 
			\toprule
			\toprule       
			\multicolumn{1}{c}{Name} 
			& \multicolumn{1}{c}{$kT_\mathrm{BB,1}$} & \multicolumn{1}{c}{$kT_\mathrm{BB,2}$} & \multicolumn{1}{c}{$\Gamma$} & \multicolumn{1}{c}{$\chi^2$/d.o.f.} \\
			\multicolumn{1}{c}{(1)} 
			& \multicolumn{1}{c}{(2)} & \multicolumn{1}{c}{(3)}
			& \multicolumn{1}{c}{(4)} & \multicolumn{1}{c}{(5)} \\
			\midrule
			PG 0052+251 & $0.12_{-0.01}^{+0.01}$ & $\dots$ & $1.88\pm0.05$ & $505/489$ \\
			\noalign{\smallskip}
			Q 0056-363 & $0.07_{-0.01}^{+0.01}$ & $0.16_{-0.01}^{+0.01}$ & $1.77\pm0.03$ & $753/764$ \\
			\noalign{\smallskip}
			Mrk 1152 & $0.15_{-0.09}^{+0.10}$ & $0.09_{-0.03}^{+0.16}$ & $1.48\pm0.03$ & $698/706$ \\
			\noalign{\smallskip}
			ESO 244-G17 & $0.10_{-0.01}^{+0.01}$ & $\dots$ & $1.68\pm0.03$ & $539/539$ \\
			\noalign{\smallskip}
			Fairall 9 & $0.07_{-0.01}^{+0.01}$ & $0.22_{-0.01}^{+0.01}$ & $1.73\pm0.01$ & $1901/1683$ \\
			\noalign{\smallskip}
			Mrk 590 & $0.11_{-0.01}^{+0.02}$ & $0.07_{-0.02}^{+0.06}$ & $1.56\pm0.02$ & $892/842$ \\
			\noalign{\smallskip}
			ESO 198-G24 & $0.06_{-0.01}^{+0.01}$ & $0.13_{-0.01}^{+0.01}$ & $1.58\pm0.01$ & $1591/1562$ \\
			\noalign{\smallskip}
			Fairall 1116 & $0.07_{-0.02}^{+0.02}$ & $0.15_{-0.01}^{+0.02}$ & $1.81\pm0.03$ & $673/633$ \\
			\noalign{\smallskip}
			1H 0419-577 & $0.06_{-0.01}^{+0.01}$ & $0.14_{-0.01}^{+0.01}$ & $1.59\pm0.01$ & $1585/1502$ \\
			\noalign{\smallskip}
			3C 120 & $0.11_{-0.01}^{+0.01}$ & $0.18_{-0.01}^{+0.01}$ & $1.67\pm0.01$ & $1780/1696$ \\
			\noalign{\smallskip}
			H 0439-272 & $0.07_{-0.01}^{+0.01}$ & $0.19_{-0.01}^{+0.01}$ & $1.70\pm0.04$ & $653/649$ \\
			\noalign{\smallskip}
			MCG-01-13-25 & $0.06_{-0.03}^{+0.04}$ & $0.17_{-0.01}^{+0.02}$ & $1.54\pm0.05$ & $477/498$ \\
			\noalign{\smallskip}
			MCG-02-14-09 & $0.08_{-0.01}^{+0.01}$ & $0.18_{-0.01}^{+0.01}$ & $1.72\pm0.02$ & $1296/1237$ \\
			\noalign{\smallskip}
			MCG+08-11-11 & $0.09_{-0.01}^{+0.01}$ & $0.19_{-0.01}^{+0.01}$ & $1.54\pm0.01$ & $1561/1578$ \\
			\noalign{\smallskip}
			PMN J0623-6436 & $0.08_{-0.02}^{+0.02}$ & $0.19_{-0.01}^{+0.02}$ & $1.60\pm0.11$ & $357/325$ \\
			\noalign{\smallskip}
			ESO 209-G12 & $0.07_{-0.01}^{+0.01}$ & $\dots$ & $1.77\pm0.03$ & $513/526$ \\
			\noalign{\smallskip}
			PG 0804+761 & $0.09_{-0.01}^{+0.01}$ & $0.18_{-0.01}^{+0.01}$ & $2.02\pm0.02$ & $776/837$ \\
			\noalign{\smallskip}
			MCG+04-22-42 & $0.08_{-0.01}^{+0.01}$ & $0.18_{-0.01}^{+0.01}$ & $1.66\pm0.04$ & $710/739$ \\
			\noalign{\smallskip}
			PG 0947+396 & $0.19_{-0.02}^{+0.02}$ & $0.08_{-0.01}^{+0.01}$ & $1.64\pm0.05$ & $393/441$ \\
			\noalign{\smallskip}
			PG 0953+414 & $0.07_{-0.01}^{+0.01}$ & $0.17_{-0.01}^{+0.01}$ & $1.91\pm0.05$ & $462/464$ \\
			\noalign{\smallskip}
			HE 1029-1401 & $0.09_{-0.01}^{+0.01}$ & $0.20_{-0.01}^{+0.01}$ & $1.83\pm0.01$ & $1405/1343$ \\
			\noalign{\smallskip}
			PG 1048+342 & $0.09_{-0.01}^{+0.01}$ & $\dots$ & $1.69\pm0.05$ & $456/468$ \\
			\noalign{\smallskip}
			PG 1115+407 & $0.08_{-0.01}^{+0.01}$ & $0.19_{-0.01}^{+0.01}$ & $1.96\pm0.09$ & $372/421$ \\
			\noalign{\smallskip}
			PG 1116+215 & $0.09_{-0.01}^{+0.01}$ & $0.19_{-0.01}^{+0.01}$ & $1.82\pm0.02$ & $932/908$ \\
			\noalign{\smallskip}
			HE 1143-1810 & $0.07_{-0.01}^{+0.01}$ & $0.15_{-0.01}^{+0.01}$ & $1.64\pm0.01$ & $1533/1467$ \\
			\noalign{\smallskip}
			PG 1202+281 & $0.11_{-0.01}^{+0.01}$ & $\dots$ & $1.82\pm0.06$ & $617/580$ \\
			\noalign{\smallskip}
			Mrk 205 & $0.07_{-0.01}^{+0.01}$ & $0.18_{-0.01}^{+0.01}$ & $2.00\pm0.01$ & $1419/1387$ \\
			\noalign{\smallskip}
			NGC 4593 & $0.18_{-0.01}^{+0.01}$ & $0.09_{-0.01}^{+0.01}$ & $1.57\pm0.02$ & $1273/1369$ \\
			\noalign{\smallskip}
			PG 1307+085 & $0.12_{-0.01}^{+0.01}$ & $\dots$ & $1.67\pm0.07$ & $365/340$ \\
			\noalign{\smallskip}
			PG 1322+659 & $0.08_{-0.01}^{+0.01}$ & $\dots$ & $2.13\pm0.06$ & $273/288$ \\
			\noalign{\smallskip}
			4U 1344-60 & $0.07_{-0.03}^{+0.04}$ & $\dots$ & $1.53\pm0.03$ & $626/655$ \\
			\noalign{\smallskip}
			Mrk 279 & $0.17_{-0.01}^{+0.01}$ & $0.08_{-0.01}^{+0.01}$ & $1.64\pm0.01$ & $1219/1306$ \\
			\noalign{\smallskip}
			PG 1352+183 & $0.08_{-0.01}^{+0.01}$ & $\dots$ & $1.92\pm0.05$ & $370/375$ \\
			\noalign{\smallskip}
			PG 1415+451 & $0.12_{-0.01}^{+0.01}$ & $\dots$ & $1.86\pm0.08$ & $522/443$ \\
			\noalign{\smallskip}
			PG 1416-129 & $0.09_{-0.01}^{+0.01}$ & $\dots$ & $1.52\pm0.03$ & $629/630$ \\
			\noalign{\smallskip}
			PG 1425+267 & $0.05_{-0.01}^{+0.01}$ & $0.22_{-0.01}^{+0.01}$ & $1.56\pm0.05$ & $634/608$ \\
			\noalign{\smallskip}
			PG 1427+480 & $0.07_{-0.01}^{+0.01}$ & $\dots$ & $1.91\pm0.03$ & $536/561$ \\
			\noalign{\smallskip}
			Mrk 841 & $0.09_{-0.01}^{+0.01}$ & $\dots$ & $1.74\pm0.02$ & $705/741$ \\
			\noalign{\smallskip}
			Mrk 290 & $0.09_{-0.01}^{+0.01}$ & $\dots$ & $1.55\pm0.03$ & $782/712$ \\
			\noalign{\smallskip}
			PG 1626+554 & $0.11_{-0.01}^{+0.01}$ & $\dots$ & $1.51\pm0.21$ & $387/395$ \\
			\noalign{\smallskip}
			PDS 456 & $0.09_{-0.01}^{+0.01}$ & $\dots$ & $2.06\pm0.01$ & $1397/1171$ \\
			\noalign{\smallskip}
			IGR J17418-1212 & $0.14_{-0.01}^{+0.01}$ & $\dots$ & $1.82\pm0.03$ & $827/908$ \\
			\noalign{\smallskip}
			Mrk 509 & $0.08_{-0.01}^{+0.01}$ & $0.16_{-0.01}^{+0.01}$ & $1.60\pm0.01$ & $2051/1799$ \\
			\bottomrule
		\end{tabular}
	\end{center}
	\begin{flushleft}
		Columns: 1 = AGN name. 2 = temperature of first blackbody in keV. 3 = temperature of second blackbody, if needed, in keV. 4 = photon index. 5 = $\chi^2$/degrees of freedom.
	\end{flushleft}
	\label{tab:4}
\end{table*}

\setcounter{table}{1}
\begin{table*}
	\caption{BLS1 spectral data (\textit{continued})}
	\begin{center}
		\begin{tabular}{lrrrr} 
			\toprule
			\toprule       
			\multicolumn{1}{c}{Name} 
			& \multicolumn{1}{c}{$kT_\mathrm{BB,1}$} & \multicolumn{1}{c}{$kT_\mathrm{BB,2}$} & \multicolumn{1}{c}{$\Gamma$} & \multicolumn{1}{c}{$\chi^2$/d.o.f.} \\
			\multicolumn{1}{c}{(1)} 
			& \multicolumn{1}{c}{(2)} & \multicolumn{1}{c}{(3)}
			& \multicolumn{1}{c}{(4)} & \multicolumn{1}{c}{(5)} \\
			\midrule
			MR 2251-178 & $0.09_{-0.01}^{+0.01}$ & $\dots$ & $1.52\pm0.01$ & $1906/1633$ \\
			\noalign{\smallskip}
			NGC 7469 & $0.09_{-0.01}^{+0.01}$ & $0.19_{-0.01}^{+0.01}$ & $1.61\pm0.01$ & $1970/1771$ \\
			\noalign{\smallskip}
			Mrk 926 & $0.10_{-0.01}^{+0.01}$ & $\dots$ & $1.59\pm0.02$ & $1040/1022$ \\
			\bottomrule
		\end{tabular}
	\end{center}
	\begin{flushleft}
		Columns: 1 = AGN name. 2 = temperature of first blackbody in keV. 3 = temperature of second blackbody, if needed, in keV. 4 = photon index. 5 = $\chi^2$/degrees of freedom.
	\end{flushleft}
	\label{tab:4_part2}
\end{table*}

\begin{table*}
	\caption{Soft excess strength for NLS1s}	
	\begin{center}
		\begin{tabular}{lccc} 
			\toprule
			\toprule       
			Name & $SX1$ & $SX2$ &  $SX3$ \\
			\midrule
			I Zw 1 & $0.30\pm0.01$ & $0.17\pm0.01$ & $(3.26\pm1.24)\times10^{-3}$ \\
			Ton S180 & $1.62\pm0.01$ & $0.70\pm0.01$ & $(8.62\pm3.10)\times10^{-3}$ \\
			Mrk 359 & $0.65\pm0.01$ & $0.26\pm0.01$ & $(1.97\pm0.75)\times10^{-3}$ \\
			Mrk 1014 & $1.02\pm0.08$ & $0.40\pm0.03$ & $(5.42\pm4.94)\times10^{-3}$ \\
			Mrk 586 & $2.10\pm0.06$ & $0.86\pm0.02$ & $(17.0\pm6.64)\times10^{-3}$ \\
			Mrk 1044 & $1.34\pm0.01$ & $0.56\pm0.01$ & $(5.78\pm1.91)\times10^{-3}$ \\
			RBS 416 & $0.85\pm0.03$ & $0.33\pm0.01$ & $(4.46\pm2.19)\times10^{-3}$ \\
			HE 0450-2958 & $1.92\pm0.07$ & $0.59\pm0.02$ & $(10.6\pm5.41)\times10^{-3}$ \\
			PKS 0558-504 & $1.35\pm0.01$ & $0.54\pm0.01$ & $(6.60\pm2.05)\times10^{-3}$ \\
			Mrk 110 & $0.63\pm0.01$ & $0.63\pm0.01$ & $(1.57\pm0.52)\times10^{-3}$ \\
			RE J1034+396 & $7.79\pm0.11$ & $3.51\pm0.05$ & $(17.3\pm11.4)\times10^{-3}$ \\
			PG 1211+143 & $0.72\pm0.01$ & $0.32\pm0.01$ & $(5.73\pm1.78)\times10^{-3}$ \\
			PG 1244+026 & $1.99\pm0.02$ & $0.95\pm0.01$ & $(6.97\pm3.00)\times10^{-3}$ \\
			IRAS 13349+2438 & $1.99\pm0.03$ & $0.42\pm0.01$ & $(7.63\pm3.21)\times10^{-3}$ \\
			PG 1402+261 & $1.66\pm0.05$ & $0.58\pm0.02$ & $(5.39\pm2.64)\times10^{-3}$ \\
			PG 1440+356 & $1.26\pm0.03$ & $0.54\pm0.01$ & $(6.16\pm2.71)\times10^{-3}$ \\
			Mrk 493 & $1.23\pm0.03$ & $0.46\pm0.01$ & $(5.16\pm1.76)\times10^{-3}$ \\
			Mrk 896 & $0.17\pm0.01$ & $0.08\pm0.01$ & $(1.46\pm0.71)\times10^{-3}$ \\
			Mrk 1513 & $0.38\pm0.02$ & $0.14\pm0.01$ & $(1.81\pm1.04)\times10^{-3}$ \\
			II Zw 177 & $0.65\pm0.03$ & $0.42\pm0.02$ & $(13.5\pm7.57)\times10^{-3}$ \\
			Ark 564 & $1.57\pm0.01$ & $0.85\pm0.01$ & $(18.4\pm5.88)\times10^{-3}$ \\
			AM 2354-304 & $0.21\pm0.04$ & $0.10\pm0.02$ & $(2.07\pm1.63)\times10^{-3}$ \\
			\bottomrule
		\end{tabular}
	\end{center}
	Note: $SX1$ = \sxone. $SX2$ = \sxtwo. $SX3$ = \sxthree.
	\label{tab:5}
\end{table*}

\begin{table*}
	\caption{Soft excess strength for BLS1s}	
	\begin{center}
		\begin{tabular}{lccc} 
			\toprule
			\toprule       
			Name & $SX1$ & $SX2$ & $SX3$ \\
			\midrule
			PG 0052+251 & $0.64\pm0.02$ & $0.24\pm0.01$ & $(3.09\pm2.08)\times10^{-3}$ \\
			Q 0056-363 & $1.09\pm0.02$ & $0.39\pm0.01$ & $(0.40\pm0.15)\times10^{-3}$ \\
			Mrk 1152 & $0.26\pm0.01$ & $0.07\pm0.01$ & $(0.45\pm0.20)\times10^{-3}$ \\
			ESO 244-G17 & $0.31\pm0.01$ & $0.12\pm0.01$ & $(0.95\pm0.69)\times10^{-3}$ \\
			Fairall 9 & $0.22\pm0.01$ & $0.08\pm0.01$ & $(2.48\pm1.66)\times10^{-3}$ \\
			Mrk 590 & $0.52\pm0.01$ & $0.18\pm0.01$ & $(1.11\pm0.45)\times10^{-3}$ \\
			ESO 198-G24 & $0.59\pm0.01$ & $0.19\pm0.01$ & $(1.34\pm0.45)\times10^{-3}$ \\
			Fairall 1116 & $0.57\pm0.05$ & $0.22\pm0.02$ & $(2.28\pm0.86)\times10^{-3}$ \\
			1H 0419-577 & $0.75\pm0.01$ & $0.26\pm0.01$ & $(2.01\pm0.67)\times10^{-3}$ \\
			3C 120 & $0.77\pm0.01$ & $0.26\pm0.01$ & $(1.21\pm0.39)\times10^{-3}$ \\
			H 0439-272 & $1.04\pm0.02$ & $0.31\pm0.01$ & $(2.42\pm1.06)\times10^{-3}$ \\
			MCG-01-13-25 & $0.82\pm0.03$ & $0.19\pm0.01$ & $(1.03\pm0.45)\times10^{-3}$ \\
			MCG-02-14-09 & $0.78\pm0.01$ & $0.25\pm0.01$ & $(1.86\pm0.62)\times10^{-3}$ \\
			MCG+08-11-11 & $0.67\pm0.01$ & $0.19\pm0.01$ & $(1.07\pm0.40)\times10^{-3}$ \\
			PMN J0623-6436 & $1.44\pm0.07$ & $0.40\pm0.02$ & $(2.61\pm2.04)\times10^{-3}$ \\
			ESO 209-G12 & $0.53\pm0.04$ & $0.21\pm0.02$ & $(2.24\pm1.56)\times10^{-3}$ \\
			PG 0804+761 & $1.14\pm0.02$ & $0.45\pm0.01$ & $(6.55\pm2.09)\times10^{-3}$ \\
			MCG+04-22-42 & $0.97\pm0.02$ & $0.28\pm0.01$ & $(1.81\pm0.68)\times10^{-3}$ \\
			PG 0947+396 & $0.88\pm0.03$ & $0.30\pm0.01$ & $(2.48\pm1.11)\times10^{-3}$ \\
			PG 0953+414 & $1.33\pm0.04$ & $0.46\pm0.01$ & $(6.90\pm2.77)\times10^{-3}$ \\
			HE 1029-1401 & $0.90\pm0.01$ & $0.31\pm0.01$ & $(3.02\pm0.96)\times10^{-3}$ \\
			PG 1048+342 & $0.47\pm0.24$ & $0.21\pm0.11$ & $(3.08\pm2.58)\times10^{-3}$ \\
			PG 1115+407 & $2.01\pm0.08$ & $0.74\pm0.03$ & $(8.47\pm3.31)\times10^{-3}$ \\
			PG 1116+215 & $1.71\pm0.03$ & $0.56\pm0.01$ & $(6.92\pm2.35)\times10^{-3}$ \\
			HE 1143-1810 & $0.87\pm0.01$ & $0.30\pm0.01$ & $(1.88\pm0.63)\times10^{-3}$ \\
			PG 1202+281 & $0.23\pm0.01$ & $0.09\pm0.01$ & $(0.71\pm0.34)\times10^{-3}$ \\
			Mrk 205 & $0.38\pm0.01$ & $0.17\pm0.01$ & $(2.95\pm0.96)\times10^{-3}$ \\
			NGC 4593 & $0.68\pm0.01$ & $0.20\pm0.01$ & $(0.34\pm0.12)\times10^{-3}$ \\
			PG 1307+085 & $0.56\pm0.04$ & $0.13\pm0.01$ & $(1.07\pm1.01)\times10^{-3}$ \\
			PG 1322+659 & $0.82\pm0.04$ & $0.50\pm0.02$ & $(10.7\pm6.95)\times10^{-3}$ \\
			4U 1344-60 & $1.90\pm1.22$ & $0.56\pm0.36$ & $(2.89\pm2.90)\times10^{-3}$ \\
			Mrk 279 & $0.85\pm0.01$ & $0.27\pm0.01$ & $(1.72\pm1.32)\times10^{-3}$ \\
			PG 1352+183 & $0.40\pm0.02$ & $0.22\pm0.01$ & $(0.38\pm0.24)\times10^{-3}$ \\
			PG 1415+451 & $1.00\pm0.03$ & $0.39\pm0.01$ & $(3.50\pm1.87)\times10^{-3}$ \\
			PG 1416-129 & $0.13\pm0.01$ & $0.04\pm0.01$ & $(0.20\pm0.09)\times10^{-3}$ \\
			PG 1425+267 & $0.92\pm0.05$ & $0.17\pm0.01$ & $(1.81\pm1.44)\times10^{-3}$ \\
			PG 1427+480 & $0.30\pm0.02$ & $0.16\pm0.01$ & $(2.07\pm1.08)\times10^{-3}$ \\
			Mrk 841 & $0.73\pm0.01$ & $0.29\pm0.01$ & $(0.52\pm0.50)\times10^{-3}$ \\
			Mrk 290 & $0.40\pm0.01$ & $0.11\pm0.01$ & $(0.70\pm0.57)\times10^{-3}$ \\
			PG 1626+554 & $0.40\pm0.03$ & $0.16\pm0.01$ & $(1.03\pm0.64)\times10^{-3}$ \\
			PDS 456 & $0.53\pm0.01$ & $0.25\pm0.01$ & $(6.08\pm2.07)\times10^{-3}$ \\
			IGR J17418-1212 & $0.28\pm0.01$ & $0.10\pm0.01$ & $(0.87\pm0.36)\times10^{-3}$ \\
			Mrk 509 & $0.84\pm0.01$ & $0.27\pm0.01$ & $(1.81\pm0.57)\times10^{-3}$ \\
			MR 2251-178 & $0.33\pm0.01$ & $0.09\pm0.01$ & $(0.66\pm0.23)\times10^{-3}$ \\
			NGC 7469 & $1.18\pm0.01$ & $0.35\pm0.01$ & $(4.57\pm1.45)\times10^{-3}$ \\
			Mrk 926 & $0.32\pm0.01$ & $0.12\pm0.01$ & $(0.96\pm0.36)\times10^{-3}$ \\
			\bottomrule
		\end{tabular}
	\end{center}
	Note: $SX1$ = \sxone. $SX2$ = \sxtwo. $SX3$ = \sxthree.
	\label{tab:6}
\end{table*}


\bsp	
\label{lastpage}
\end{document}